\documentclass[sigconf]{acmart}
\usepackage{weiwAlgorithm}
\usepackage{makecell}
\AtBeginDocument{%
  \providecommand\BibTeX{{%
    Bib\TeX}}}

\copyrightyear{2026}
\acmYear{2026}
\setcopyright{cc}
\setcctype{by}
\acmConference[KDD '26]{Proceedings of the 32nd ACM SIGKDD Conference on Knowledge Discovery and Data Mining V.2}{August 09--13, 2026}{Jeju Island, Republic of Korea}
\acmBooktitle{Proceedings of the 32nd ACM SIGKDD Conference on Knowledge Discovery and Data Mining V.2 (KDD '26), August 09--13, 2026, Jeju Island, Republic of Korea}

\newcommand{\eat}[1]{}

\usepackage{amsthm}
\usepackage{graphicx}
\usepackage{balance}  
\usepackage{lipsum}
\usepackage{amsmath}
\usepackage{physics}
\usepackage{subfigure}
\usepackage{tabularx}
\usepackage{color}
\usepackage{colortbl}
\usepackage{float}
\usepackage{listings}
\usepackage{multirow}
\usepackage[normalem]{ulem}
 \usepackage{longtable}
\usepackage{enumitem}
\usepackage{multirow}
\usepackage{booktabs}
\usepackage{capt-of}
\usepackage{pifont}
\usepackage{ulem}
\usepackage{soul}
\usepackage{cancel}
\usepackage{bm}
\usepackage{url}
\usepackage{caption}

\def\BibTeX{{\rm B\kern-.05em{\sc i\kern-.025em b}\kern-.08em
    T\kern-.1667em\lower.7ex\hbox{E}\kern-.125emX}}

 \graphicspath{{./Graph/}, {./Fig/}, {./Legend/}}

\long\def\comment#1{}

\newcounter{example}[section]
\renewcommand{\theexample}{\nthesection.\arabic{example}}
\newenvironment{example}{
     \refstepcounter{example}
     {\vspace{1ex} \noindent\bf  Example  \theexample:}}{
     \vspace{1ex}} 

\newcounter{definition}[section]
\renewcommand{\thedefinition}{\nthesection.\arabic{definition}}
\newenvironment{definition}{
     \refstepcounter{definition}
     {\vspace{1ex} \noindent\bf  Definition  \thedefinition:}}{
     \vspace{1ex}} 

\newcounter{theorem}[section]
\renewcommand{\thetheorem}{\nthesection.\arabic{theorem}}
\newenvironment{theorem}{\begin{em}
        \refstepcounter{theorem}
        {\vspace{1ex} \noindent\bf  Theorem  \thetheorem:}}{
        \end{em}\vspace{1ex}} 

\newcounter{lemma}[section]
\renewcommand{\thelemma}{\nthesection.\arabic{lemma}}
\newenvironment{lemma}{\begin{em}
        \refstepcounter{lemma}
        {\vspace{1ex}\noindent\bf Lemma \thelemma:}}{
        \end{em}\vspace{1ex}} 

\newcounter{corollary}[section]
\renewcommand{\thecorollary}{\nthesection.\arabic{corollary}}

\newcounter{proposition}[section]
\renewcommand{\theproposition}{\nthesection.\arabic{proposition}}

\newcounter{remark}[section]
\renewcommand{\theremark}{\nthesection.\arabic{remark}}

\newcommand{\proofsketch}{\noindent{\bf Proof Sketch: }}

\newcommand{\nthesection}{\arabic{section}}

\newcommand{\eop}{\hspace*{\fill}\mbox{$\Box$}\vspace*{1ex}}

\newcommand{\stitle}[1]{\noindent{\bf #1}}

\newcommand{\green}[1]{\textcolor{green}{}}

\newcommand{\kw}[1]{{\ensuremath {\mathsf{#1}}}\xspace}

\newcommand\bigutimes{\mathop{\ooalign{$\bigcup$\cr%
   \hfil\raise0.36ex\hbox{$\scriptscriptstyle\boldsymbol{\times}$}\hfil\cr}}}
\newcommand\bigumius{\mathop{\ooalign{$\bigcup$\cr%
   \hfil\raise0.36ex\hbox{$\scriptscriptstyle\boldsymbol{-}$}\hfil\cr}}}

\newcommand{\deepcolor}{\cellcolor[rgb]{0.8745, 0.8235, 0.7137}}
\newcommand{\lightcolor}{ \cellcolor[rgb]{0.9882,0.9373,0.8196}}

\begin{document}

\title{Generalized Range Filtering Approximate Nearest Neighbor Search: Containment and Overlap [Technical Report]}


\author{Yingfan Liu}
\affiliation{%
  \institution{Xidian University}
  \city{Xi'an}
  \country{China}
}
\email{liuyingfan@xidian.edu.cn}

\author{Tong Wu}
\affiliation{%
  \institution{Xidian University}
  \city{Xi'an}
  \country{China}
}
\email{twu_1@stu.xidian.edu.cn}

\author{Jiadong Xie}
\authornote{Jiadong Xie and Jiangtao Cui are the corresponding authors.}
\affiliation{%
  \institution{The Chinese University of Hong Kong}
  \city{Hong Kong SAR}
  \country{China}
}
\email{jdxie@se.cuhk.edu.hk}

\author{Yang Zhao}
\affiliation{%
  \institution{Xidian University}
  \city{Xi'an}
  \country{China}
}
\email{zhaoyang_1@stu.xidian.edu.cn}

\author{Jeffrey Xu Yu}
\affiliation{%
  \institution{The Hong Kong University of Science and Technology (Guangzhou)}
  \city{Guangzhou}
  \country{China}
}
\email{jeffreyxuyu@hkust-gz.edu.cn}

\author{Jiangtao Cui}
\authornotemark[1]
\affiliation{%
  \institution{Xi'an University of Posts and Telecommunications‌}
  \city{Xi'an}
  \country{China}
}
\affiliation{%
  \institution{Xidian University}
  \city{Xi'an}
  \country{China}
}
\email{cuijt@xidian.edu.cn}




\begin{abstract}
Approximate nearest neighbor (ANN) search with range filters has recently garnered significant attention. This paper delves into a generalized form of this problem, i.e., ANN search with exact range-range (RR) predicates on a range-valued attribute, named RR filtering ANN (RRANN). Specifically, given $n$ vectors in $\mathbb{R}^d$, each vector $v_i$ is associated with a numeric range $[l_i, r_i]$, symbolizing aspects like a price range or time interval. An RRANN query $(v_q, l_q, r_q)$ aims at finding $k$ vectors closest to $v_q$ within the vectors satisfying an arbitrary RR predicate defined between the query range $[l_q, r_q]$ and the object range $[l_i, r_i]$. The RR predicate remains unspecified, enabling user-defined conditions. It may encompass containment ($[l_i, r_i] \subseteq [l_q, r_q]$ or $[l_q, r_q] \subseteq [l_i, r_i]$), overlap ($l_i \le l_q \le r_i \le r_q$ or $l_q \le l_i \le r_q \le r_i$), or a disjunction of them. RRANN has broad applications in queries related to price ranges or time intervals, and it generalizes existing variants of ANN search with range filters. However, existing dedicated approaches for these problems lack the capacity to support queries with arbitrary RR predicates. Hence, we introduce a new approach, labeled multi-segment tree graph. It efficiently handles arbitrary RR predicates by avoiding traversal through non-predicate-satisfied nodes, and keeps equivalent index size and construction time to state-of-the-art methods for RFANN. Extensive experiments on real-world data demonstrate the efficacy of our approach in RRANN queries, achieving up to 12.5x speedups with the same accuracy as the baselines. Moreover, our approach attains comparable RFANN search performance and notably superior IFANN and TSANN search performance compared to the respective state-of-the-art approaches. Our code is available at \url{https://github.com/FanEDG/MSTG}.
\end{abstract}

\maketitle

\section{Introduction}


With the recent advancements in embedding models leveraging machine learning techniques, diverse objects, such as images~\cite{nasrabadi1988image} and texts~\cite{word2vec}, are embedded into high-dimensional vectors to capture their semantic information. This has led to the rise in popularity of vector databases in both research communities and the industry~\cite{abs-2501-11216,PanWL24,AziziEP23,abs-2504-10326,ZhuCGMZZ24}. 
In vector databases, the fundamental operation is the approximate $k$-nearest neighbor search ($k$-ANNS), which retrieves $k$ vectors sufficiently close to a given query vector. 
Vector databases like Milvus~\cite{milvus} and AnalyticDB~\cite{ADBV} enhance search precision by integrating $k$-ANN search with attribute-based filters.


{
In this paper, we study the $k$-ANNS with range filters.
In numerous applications, each object in the dataset $O$ is represented as $o_i=(v_i, l_i, r_i)$, where $v_i \in \mathbb{R}^d$ is a $d$-dimensional vector and $[l_i, r_i] \subset \mathbb{R}$ is a numeric range. Similarly, each query $q$ consists of a vector $v_q$ and a range $[l_q, r_q]$.
Here, the range-range (RR) predicates between the query range and the object range are not specified, allowing user-defined conditions. 
As illustrated in Fig.~\ref{fig:illus_rf}, there are four atomic conditions, including two types of containment and two types of overlap: \textcircled{1} query left-overlap: $l_i\leq l_q\leq r_i\leq r_q$; \textcircled{2} query-contained: $l_i\leq l_q\leq r_q\leq r_i$; \textcircled{3} query right-overlap: $l_q\leq l_i\leq r_q\leq r_i$; and \textcircled{4} query-containing: $l_q\leq l_i\leq r_i\leq r_q$.
The RR predicates can be one of them, or a disjunction of them.
This problem, $k$-ANNS with arbitrary RR predicates, referred to as \textit{range-range filtering $k$-ANN} (RRANN), has broad applications.


\stitle{Price Range Querying:} 
The price of an object in real-world scenarios can be a continuous interval, such as a stock price range or prices of products on comparison shopping websites that aggregate product data across multiple online retailers. By using RRANN queries, users can search for objects that meet specific price predicates. For example, a user might seek a shoe resembling a provided image on comparison shopping websites, priced between \$50 and \$100. 
RRANN can be utilized to retrieve products $(v_i,l_i,r_i)$ whose image vectors have the smallest distances to the query image vector, and simultaneously satisfy $[l_i,r_i]\cap [50,100]\neq \emptyset$.

\stitle{Time-Relevant Querying:}
Consider multiple traffic cameras monitoring cars passing on a state highway. Each camera detects and extracts a feature vector representing every car in the video stream. These feature vectors, along with their time ranges on the highway, are stored in a database. 
Upon receiving a query containing a particular car image and a given time interval, RRANN can locate similar cars that traversed the highway within the query time range.

\begin{figure}[t]
\centering
\includegraphics[width=\linewidth]{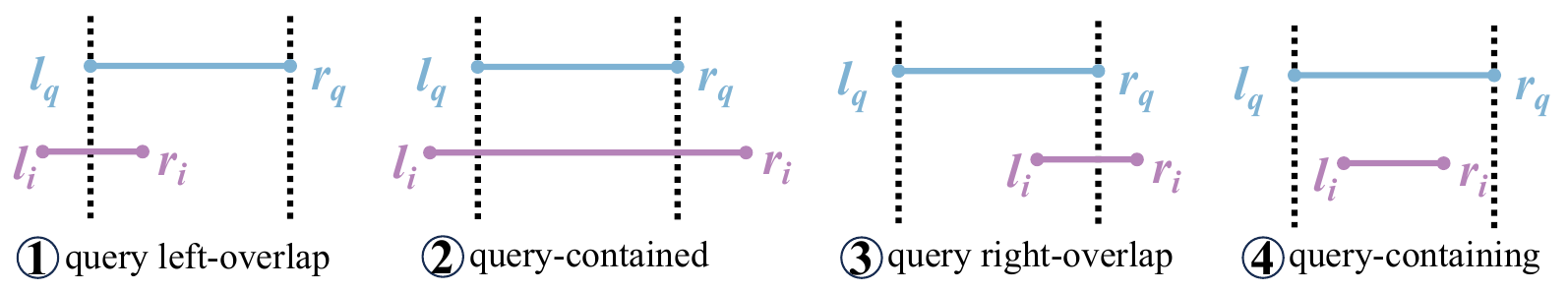}
\vspace{-5mm}
\caption{The atomic conditions of the RR predicates}
\label{fig:illus_rf}
\vspace{-3mm}
\end{figure}

\begin{table*}[t]
    \centering
     \caption{Comparisons between RRANN with other range-filtering $k$-ANN problems}
     \vspace{-3mm}
    \resizebox{0.95\linewidth}{!}{
    \begin{tabular}{|c|c|c|c|c|c|c|c|}
    \hline
\multicolumn{1}{|c|}{\multirow{2}{*}{\textbf{Problems}}} & \multicolumn{1}{c|}{\multirow{2}{*}{\textbf{Object Attribute}}} & \multicolumn{1}{c|}{\multirow{2}{*}{\textbf{Query Attribute}}} & \multirow{2}{*}{\textbf{RR Predicate}} & \multicolumn{4}{c|}{\textbf{QPS at Recall@10=0.95 on Gist dataset when selectivity is $10\%$}} \\ \cline{5-8}
&&&    & iRangeGraph~\cite{iRange}    &     Hi-PNG~\cite{Hi-PNG}   &   TS-Graph~\cite{TSANN}      & MSTG (ours)      \\ \cline{2-3}
         \hline \hline
     RFANN~\cite{iRange} & point-valued ($l_i = r_i$) & \deepcolor range-valued & $[l_i,r_i] \subseteq [l_q, r_q]$ (\textcircled{4})&{\lightcolor 752.997}&{127.238}&-&{\deepcolor 753.215} \\
         \hline
         IFANN~\cite{Hi-PNG} & \deepcolor range-valued & \deepcolor range-valued & $[l_i, r_i] \subseteq [l_q, r_q]$ (\textcircled{4})&-&{\lightcolor 156.790}& - &{\deepcolor 873.672} \\
         \hline
         TSANN~\cite{TSANN} & \deepcolor range-valued & point-valued ($l_q = r_q$) & $[l_q,r_q] \subseteq [l_i, r_i]$ (\textcircled{2})&-&-&{\lightcolor 238.851}&{\deepcolor 925.845} \\
         \hline
         RRANN (ours) & \deepcolor range-valued & \deepcolor range-valued & \deepcolor arbitrary&{-}&-&{-}&{\deepcolor 604.572}\\
         \hline
    \end{tabular}
    }
    \label{tb:cmp_problems}
    \vspace{-3mm}
\end{table*}

\stitle{General Form of other $k$-ANNS with Range Filters:}
{
There are three variations of $k$-ANNS with range filters: (1) range-filtering $k$-ANN (RFANN)~\cite{iRange,SeRF,vbase,ADBV,ACORN,UNIFY,WST,JiangYZHSZLW25,XieYTL25} with point-valued object attribute and range-valued query attribute, 
(2) interval-filtering $k$-ANN (IFANN)~\cite{Hi-PNG} with range-valued object attribute and range-valued query attribute but limited RR predicates,
and (3) timestamp $k$-ANN (TSANN)~\cite{TSANN} with range-valued object attribute and point-valued query attribute.
These variations, as outlined in Table~\ref{tb:cmp_problems}, are all special cases of our RRANN problem. Specifically, the RR predicate of RFANN is $[l_i,r_i]\subseteq [l_q,r_q]$ (i.e., case \textcircled{4}) with $l_i=r_i$; the RR predicate of IFANN is $[l_i,r_i]\subseteq [l_q,r_q]$ (i.e., case \textcircled{4}); the RR predicate of TSANN is $[l_q,r_q]\subseteq [l_i,r_i]$ (i.e., case \textcircled{2}) with $l_q=r_q$.
}

The ideal solution to $k$-ANNS with filters is to conduct a $k$-ANNS on a proximity graph (PG) pre-built for the subset $O[R_q] \subseteq O$ just satisfying the query predicate $R_q$, since PGs are recognized as the state-of-the-art (SOTA) methods for $k$-ANNS~\cite{Dpg, survey2021, AziziEP25}.
Thus, recent works focus on designing a dedicated index atop PGs for $k$-ANNS with a specific range filter~\cite{iRange,TSANN,Hi-PNG,UNIFY,JiangYZHSZLW25},
such that the index can swiftly extract a PG containing vectors in $O[R_q]$ for search.
Among them, iRangeGraph~\cite{iRange}, Hi-PNG~\cite{Hi-PNG}, and TS-Graph~\cite{TSANN} are SOTA approaches for RFANN, IFANN, and TSANN, respectively.
To answer RFANN queries, iRangeGraph~\cite{iRange} utilizes a segment tree to partition objects based on their numeric attributes and constructs a PG for each segment tree node. For any given query range, iRangeGraph ensures that at most $O(\log n)$ pre-built PGs are needed to be merged to form the PG on $O[R_q]$ for search.
For IFANN queries, Hi-PNG~\cite{Hi-PNG} transfers each object range to a 2D point, and then builds a QuadTree on 2D space with a PG on each tree node.
During search, it rapidly identifies the tree nodes concerning $R_q$ and then merges the results of $k$-ANNS on those nodes to return.
For TSANN queries, TS-Graph~\cite{TSANN} constructs and compresses a series of PGs of each discrete timestamp to efficiently identify neighbors that satisfy $R_q$ during search.
However, as discussed in Section~\ref{sec:limit}, these methods fail in extending support to arbitrary RR predicates. It is because their index cannot efficiently extract a PG exactly containing objects in $O[R_q]$.
Table~\ref{tb:cmp_problems} demonstrates the search performance of each problem's SOTA approach, where ``-'' denotes that they cannot be extended to answer these queries.
Existing approaches struggle to extend their solutions or exhibit poor performance when utilized to solve other problems, while ours exhibit the best performance across all problems.
In this paper, we aim to design a novel dedicated index for efficient RRANN search.
Our main contributions are summarized below.
\ding{202} We first solve RRANN with a query-contained filter, i.e., $l_i\le l_q \le r_q \le r_i$. 
For $l_i$ values, objects satisfying $l_i\le l_q$ form prefixes of a sorted sequence $L=\{l_i | o_i \in O\}$ in ascending order of $l_i$. We propose an index called multi-segment tree graph (MSTG), which constructs $|L|$ segment trees of different sequence prefixes, with each segment tree built based on the $r_i$ values to handle the condition $r_i\ge r_q$. 
For each tree node of MSTG, we build a PG.
Hence, RRANN with a query-contained filter can be processed using a PG merged from $O(\log n)$ PGs derived from nodes in one of the $|L|$ segment trees in MSTG. 
\ding{203} 
We further enhance the efficiency of MSTG in both building time and index size. 
To achieve this, we introduce a labeled MSTG with an incremental construction method to avoid the repeated computations and merge the same edge on multiple PGs with labels for lossless compression.
\ding{204} We extend MSTG from addressing RRANN with a query-contained filter to handling RRANN with arbitrary RR predicates via simple modifications.
\ding{205}
Extensive experiments demonstrate the effectiveness of MSTG in RRANN and its variants: RFANN, TSANN, and IFANN. MSTG surpasses baselines on RRANN queries {by up to $12.5$x on efficiency while achieving the same recall.}
Moreover, compared with the SOTA approaches, our approach has comparable performance on RFANN queries and significantly improved recall and efficiency on TSANN and IFANN queries.
}

\section{Preliminaries}
\label{sec:preli}


Let $D\subset \mathbb{R}^d$ be a dataset with $n$ $d$-dimensional vectors. For any two vectors $u,v\in \mathbb{R}^d$, let $\delta(u, v)$ denote the distance between two vectors, and the L2 norm (i.e., Euclidean distance) is used by default in this work. We first define the $k$-ANN problem.

\stitle{$k$-ANN Problem}:
 Given a dataset $D \subset \mathbb{R}^d$ and a query $q \in \mathbb{R}^d$, $k$-ANN query returns $k$ vectors in $D$ that are sufficiently close to $q$. 


{
In this paper, we focus on $k$-ANN search with a RR predicate on a single range-valued attribute. 
To be specific, let $A=\{a_1,a_2,\cdots,a_{|A|}\}$ be the numeric attribute (e.g., prices, timestamps), whose domain $Dom(A)$ has a \textit{total order}, i.e., assuming that $a_1<a_2<\cdots<a_{|A|}$. Each vector $v_i\in D$ is associated with an interval attribute on $Dom(A)$, denoted as $[l_i, r_i]$, where $l_i,r_i\in Dom(A)$ and $l_i\le r_i$, i.e., $[l_i,r_i]\subseteq Dom(A)$. We consider a dataset $O$ of $n$ objects, and define each object $o_i \in O$ ($1 \leq i \leq n$) as $o_i = (v_i, l_i, r_i)$, where $v_i \in \mathbb{R}^d$ and $[l_i,r_i]\subseteq Dom(A)$.
For a query $q = (v_q, l_q, r_q)$, the query vector $v_q \in \mathbb{R}^d$ is associated with a numeric range $[l_q,r_q]\subseteq Dom(A)$. 
Let $R_q$ be the query RR predicate, and $O[R_q] = \{o_i \in O | R_q([l_i, r_i], [l_q, r_q] = \texttt{true})\}$ the set of objects that satisfy $R_q$. 
}

{
Now, let us consider the potential forms of the predicate $R_q$. According to Allen’s Interval Algebra~\cite{Allen83},
there are a total of 13 base relations between two ranges. Among them, 11 relations could be reduced to four atomic \textbf{range-range (RR) predicates} as shown in Fig.~\ref{fig:illus_rf}, while the remaining two could also be supported by our method, as discussed in Appendix~\ref{sec:correspondence}. 
Specifically, four atomic cases are defined as:
\textcircled{1} query left-overlap: $l_i\leq l_q\leq r_i\leq r_q$;
\textcircled{2} query-contained: $l_i\leq l_q\leq r_q\leq r_i$;
\textcircled{3} query right-overlap: $l_q\leq l_i\leq r_q\leq r_i$;
and \textcircled{4} query-containing: $l_q\leq l_i\leq r_i\leq r_q$.
}
In this work, $R_q$ can be defined as one of the four cases or a disjunctive combination thereof. For instance, $R_q$ could be set as \textcircled{2} to ensure that the ranges of the qualified objects fully cover the query range.
Alternatively, setting $R_q$ as \textcircled{1}$\vee$\textcircled{2}$\vee$\textcircled{3}$\vee$\textcircled{4} 
indicates $[l_i,r_i]\cap [l_q,r_q]\neq \emptyset$. 
Based on this, we define our problem.

\begin{definition}[Range-Range Filtering $k$-ANN (RRANN)]
\label{def:rrann}
Given an object set $O$, a query $q=(v_q,l_q,r_q)$, and a RR predicate $R_q$, RRANN query aims to return $k$-ANN of $v_q$ within set $O[R_q]$.
\end{definition}






\emph{To the best of our knowledge, this work is the first attempt to study $k$-ANN with the general form of RR predicates.}
As depicted in Table~\ref{tb:cmp_problems}, we introduce the existing variations of $k$-ANN search with RR predicates, which are all special cases of our RRANN problem.


\stitle{Range-Filtering $k$-ANN (RFANN)~\cite{SeRF,iRange,JiangYZHSZLW25,UNIFY}:} It considers each $v_i$ is associated with a numerical value $a_i$, and a query $(v_q,l_q,r_q)$ aims to find $k$-ANN satisfying $a_i\in [l_q,r_q]$.
RRANN will transform to RFANN by adding one constraint on object attributes $l_i=r_i$, and a specific RR predicate, i.e., let $R_q$ be the atomic case \textcircled{4}.

\stitle{Interval-Filtering $k$-ANN (IFANN)~\cite{Hi-PNG}:} It considers each $v_i$ has an interval $[l_i,r_i]$, and for a query $(v_q,l_q,r_q)$, it aims to find $k$-ANN satisfying $[l_i,r_i]\subseteq [l_q,r_q]$. RRANN will transform to IFANN by specifying the RR predicate $R_q$ as atomic case \textcircled{4}.

\stitle{Timestamp $k$-ANN (TSANN)~\cite{TSANN}:} It considers each $v_i$ has a time interval $[l_i,r_i]$. Given a query vector $v_q$ and timestamp $t_q$, it aims to find $k$-ANN satisfying $t_q\in [l_i,r_i]$.
RRANN can transform to TSANN by limiting $l_q=r_q$ and let RR predicate be the case \textcircled{2}.


Next, we brief the SOTA $k$-ANNS method, i.e., proximity graph.

\stitle{Proximity Graphs (PG)}:
PGs, such as HNSW \cite{Hnsw}, NSG \cite{Nsg}, $\tau$-MNG~\cite{taumg} and ALMG~\cite{ALMG}, have been recognized as the SOTA $k$-ANNS approaches according to several recent studies~\cite{survey2021,FastPG,AziziEP25}.
Let $G=(V, E)$ be a PG defined over a set $D \subset \mathbb{R}^d$ of vectors, where $V$ is its vertex set and $E$ is its edge set. Each vertex $u \in V$ uniquely represents a vector in $D$, and $(u, v) \in E$ indicates that $v$ is a close neighbor of $u$ in the vector space. 
We use $N_G(u)$ to denote the neighbors of $u$ in the PG $G$, i.e., $N_G(u)=\{v\in V\mid (u,v)\in E\}$.
Different graphs share the same vertex set but distinct edge sets due to their specific edge selection strategies that prune redundant neighbors over a set of close neighbors for each vector. 
{Despite variations in graph structures, existing PGs share a common $k$-ANN search algorithm~\cite{Nsg,FastPG}, which employs a greedy approach that progressively approaches the nodes that are closest to the query. 
The details of the search procedure on a PG are included in Appendix~\ref{sec:details_alg} (Algorithm~\ref{alg:knn_search}).}
In this paper, \emph{we employ HNSW~\cite{Hnsw} as the default PG}, which is one of the SOTA methods and naturally supports the insertions of new vectors.
Furthermore, we present a summary of notations in Table~\ref{tb:notations} to enhance the readability.


\begin{table}[t]
\centering
\caption{Summary of Notations}
\vspace{-3mm}
\label{tb:notations}
\resizebox{0.85\linewidth}{!}{
\begin{tabular}{|l|l|}
\hline
\textbf{Notation} & \textbf{Definition} \\
\hline \hline
$A \subset \mathbb{R}$ & the domain of the numeric attribute \\  \hline
$a_i \in A$ & an attribute value \\  \hline
$o_i$ & an object \\  \hline
$v_i$ & the vector of object $o_i$ \\  \hline
$v_q$ & the query vector \\  \hline
$d$ & the vector dimensionality \\  \hline
$\delta(u,v)$ & the distance between two vectors \\  \hline
$[l_i, r_i]$ & the range of object $o_i$ \\  \hline
$[l_q, r_q]$ & the query range \\ \hline
$R_q$ & the RR predicate specified by query $q$ \\ \hline
$O_{R_q}$ & the set of objects satisfying predicate $R_q$ \\ \hline
$G=(V,E)$ & a PG $G$ with vertex set $V$ and edge set $E$ \\ \hline
$N_G(u)$ & the neighbors of $u$ in the graph $G$ \\ \hline
$a_x$ & the $x$-th smallest attribute value in $A$ \\ \hline
$O_x$ & the set of objects whose ranges satisfy $l_i \le a_x$ \\ \hline
$\mathcal{T}_x$ & the segment tree that manages $O_x$ \\ \hline
$G_x$ & the segment tree graph that manages $O_x$ \\ \hline
\end{tabular}
}
\vspace{-3mm}
\end{table}

\section{Limitations of Existing Approaches}\label{sec:limit}

We review the existing approaches to $k$-ANNS with filters, and contemplate their potential to address our problem while also identifying their limitations. 
Those methods could be divided into two categories, i.e., (1) the general-purpose approaches for arbitrary filters, and (2) the dedicated methods for a specific filter. 


\vspace{1mm}
\stitle{General-Purpose Approaches:}
These approaches can support $k$-ANNS with arbitrary filters, including pre-filtering~\cite{ADBV,milvus,vbase}, post-filtering~\cite{ADBV,milvus}, Milvus~\cite{milvus}, VBASE~\cite{vbase}, and ACORN~\cite{ACORN}.
Detailed discussions on them could be found in Appendix~\ref{sec:details_gpa}. Here, we focus on their issues.
\noindent\underline{\textbf{Issues:}}
Although supporting $k$-ANNS with arbitrary filters, they exhibit suboptimal performance, as shown in Section~\ref{sec:exp}, because they fail to avoid verifying vectors that do not satisfy the query predicate, due to the general-purpose index. Unfortunately, each vector verification requires an expensive distance computation, making these methods inefficient in practice.

\vspace{1mm}
\stitle{Dedicated Indexes for Range Filters:}
There are several dedicated approaches designed for $k$-ANNS with range filters.
To be specific, iRangeGraph~\cite{iRange} is the SOTA method among existing RFANN approaches~\cite{UNIFY,SeRF,JiangYZHSZLW25,XieYTL25}, Hi-PNG~\cite{Hi-PNG} is proposed for IFANN, and TS-Graph~\cite{TSANN} is designed for TSANN. 
Their key idea is to pre-build a series of PGs for some attribute ranges, which can help to efficiently online form a PG $G^{\prime}$ exactly containing objects satisfying the query predicate. Next, the $k$-ANNS with range filters transfers to $k$-ANNS on $G^{\prime}$, which could be efficiently answered by Algorithm~\ref{alg:knn_search}.




\eat{
\stitle{SeRF~\cite{SeRF}}:
As the first dedicated RFANN method, it creates a new graph structure, 1D Segment Graph (1DSG), to address the half-bounded range filter, $[l_q, r_q=\infty)$. SeRF builds $|A|$ HNSW graphs $G_1, G_2, \ldots, G_{|A|}$, where each $G_x$ manages the vectors with attributes $\geq a_x\in A$.
Given a query range $[l_q, \infty)$, $G_y$ is identified to support the RFANN query by $k$-ANN search on $G_y$, where {$a_{y+1} < l_q \le a_y$} (assuming $a_0=+\infty$).
For arbitrary query ranges, SeRF introduces 2D Segment Graph (2DSG) by constructing $|A|$ 1DSGs $\{\mathcal{G}_1, \ldots, \mathcal{G}_{|A|}\}$ and then compressing them. Each 1DSG $\mathcal{G}_x$ manages objects with attributes $\leq a_x$. Given a query range $[l_q, r_q]$, SeRF identifies the 1DSG $\mathcal{G}_y$ such that {$a_y \leq r_q < a_{y+1}$} (assuming $a_{|A|+1}=+\infty$), enabling a search on $\mathcal{G}_y$ with the query range transformed into $[l_q, +\infty)$. 
However, SeRF suffers from two issues, i.e. (1) excessive index size due to compressing $|A|$ 1DSG, and (2) poor RFANN performance caused by its skipping the construction of certain 1DSGs for efficiency in index size and building cost, which leads to verifications of out-of-range vectors during search. 
}

{
\noindent\underline{\textbf{iRangeGraph~\cite{iRange}:}}
It employs a segment tree to organize the objects with $a_i$ as the key, where each tree node contains a subset of objects rooted at itself and the root contains all the objects. 
Hence, each object appears in $O(\log n)$ tree nodes. 
For each tree node, it builds a PG, called an elemental graph, and thus each object has $O(\log n)$ neighbor sets from different elemental graphs. 
Given $[l_q, r_q]$, let $G^{\prime} = (V^{\prime}, E^{\prime})$ be the dedicated PG for the in-range objects, which is built online and virtually. 
It retrieves at most $O(\log n)$ PGs covering $[l_q, r_q]$ to build $G^{\prime}$ by merging them.
It limits the out-degree of each $u\in V^{\prime}$ to a threshold $m$ by a high-layer-first pruning.
Finally, $k$-ANNS on $G^{\prime}$ returns the result of an RFANN query.  
}

\noindent\underline{\textbf{Hi-PNG~\cite{Hi-PNG}:}}
Hi-PNG is the only IFANN approach.
It treats each object range $[l_i, r_i]$ as a point $(l_i, r_i)$ in $\mathbb{R}^2$ and the query range $[l_q, r_q]$ as a rectangle $[l_q, r_q] \times [l_q, r_q] \subseteq \mathbb{R}^2$.
Hence, $[l_i, r_i] \subseteq [l_q, r_q]\Leftrightarrow (l_i, r_i) \in [l_q, r_q] \times [l_q, r_q]$.
In this way, the filter is transformed into finding the points within the query rectangle. 
Thus, Hi-PNG employs a QuadTree~\cite{samet1984quadtree} to manage those points in $\mathbb{R}^2$ and build a PG for each tree node.
During the search process, it finds a minimum set of tree nodes intersecting with the query rectangle and then returns the merged results, each of which is obtained by $k$-ANNS or post-filtering on the corresponding PG.

\noindent\underline{\textbf{TS-Graph~\cite{TSANN}:}}
TS-Graph is the only TSANN method.
It is based on the idea that builds $|A|$ PGs $G_1, \ldots, G_{|A|}$, where $G_i$ manages all the objects with the ranges containing $a_i\in A$. Here, $A$ indicates the set of timestamps. 
Next, it compresses those graphs into a single index by merging the repeated nodes and edges. For a TSANN query $(v_q\in \mathbb{R}^d, t_q \in A)$, it extracts $G_{t_q}$ from the compressed graph, and then conducts $k$-ANNS on $G_{t_q}$ as the query result.

\noindent\underline{\textbf{Issues:}}
When attempting to adapt existing dedicated approaches to address the RRANN problem, inherent issues become apparent. As follows, we meticulously analyze these approaches individually.
%
%
%
\ding{202}
\stitle{iRangeGraph:} 
To enable iRangeGraph to support each object with a numerical range, we consider dividing each numerical range into multiple numerical values.
Specifically, we can divide each $[l_i, r_i]$ into numerical values, i.e., assuming each $o_i$ has a numerical set $\mathcal{I}_i=A \cap [ l_i, r_i]$. Hence, a PG of a segment tree node representing the range $[l,r]$ will contain the object $o_i=(v_i,\mathcal{I}_i)$ if $[l,r]\cap \mathcal{I}_i\neq \emptyset$.
Like iRangeGraph, given a query range $(l_q, r_q)$, we consider online forming a PG $G^{\prime}$ containing objects that satisfy the RR predicate to transform the problem into $k$-ANNS on $G^{\prime}$.
%
\noindent{\underline{\textbf{Issues}}:}
First, unlike the original iRangeGraph where each object appears in at most $O(\log n)$ tree nodes, each object appears in at most $O(\log n\cdot |\mathcal{I}_i|)$ tree nodes to deal with for RRANN queries, which significantly increases the index size and building time.
Second, even with such a heavy index, it is still impossible to extract a PG that exactly contains the objects satisfying the arbitrary RR predicate. It is because the range of each object has been divided into multiple numerical values, which cannot process complex constraints on a range. 
For example, consider the atomic case \textcircled{1} where $l_i\le l_q$ and $l_q\leq r_i\leq r_q$. The PGs extracted from the iRangeGraph contain objects in $\{o_i=(v_i,\mathcal{I}_i)\mid \mathcal{I}_i\cap[l_q,r_q]\neq \emptyset\}$, which cannot ensure $l_q \le r_i\le r_q$ nor $l_i\le l_q$.
\ding{203}
\stitle{Hi-PNG:} 
We consider directly utilizing Hi-PNG to support RRANN queries, since the object and query attributes of IFANN are the same as RRANN.
%
\noindent{\underline{\textbf{Issues}}:}
Hi-PNG cannot support arbitrary RR predicates since they cannot always transform into the point-in-rectangle query $(l_i, r_i) \in [l_q, r_q] \times [l_q, r_q]$.
For example, considering the RR predicate \textcircled{1}$\vee$\textcircled{3} where $l_i\le l_q\leq r_i\leq r_q$ or $l_q\le l_i\le r_q\le r_i$, it cannot form a rectangle area.
Furthermore, Hi-PNG needs $k$-ANNS on multiple PGs and extra post-filtering operations on the results of some PGs, even within the IFANN predicate $[l_i, r_i] \subseteq [l_q, r_q]$. This results in the traversal of unnecessary nodes and their distance computations during the search process, ultimately compromising search efficiency.
Exp. 4 (Fig.~\ref{fig:ifann}) demonstrates its poor search performance, with low QPS and failing to achieve high recall.
%
\ding{204}
\stitle{TS-Graph:}
To enable TS-Graph support RR predicates, we consider dividing the query range $[l_q, r_q]$ into a set $\mathbf{I}_q=A \cap [l_q,r_q]$. For querying, the results are merged from  $|\mathbf{I}_q|$ TSANN separate queries, i.e., $(v_q, t_j)$ for each $t_j \in \mathbf{I}_q$.
\noindent{\underline{\textbf{Issues}}:}
First, this approach cannot support arbitrary RR predicates. Since TS-Graph finds objects satisfying $t_q \in [l_i, r_i]$, $|\mathbf{I}_q|$ TSANN separate queries will find objects $\{o_i=(v_i,l_i,r_i)\mid [l_q,r_q] \cap [l_i, r_i]\neq \emptyset\}$. It is just the disjunction of all four atomic cases instead of an arbitrary RR predicate.
Second, for an RRANN query, TS-Graph needs $|\mathbf{I}_q|$ queries, which suffers from a significant inefficiency issue, especially when $|\mathbf{I}_q|$ is big. Besides, TS-Graph cannot achieve high recall, as shown in Exp. 5 (Fig.~\ref{fig:tsann}).
\section{Our Method}
\label{sec:pstg}

As discussed above, existing dedicated approaches do not support RRANN in an ideal manner. Hence, we aim to design a new index that can efficiently support RRANN with arbitrary RR predicates, ensuring that non-satisfying objects are bypassed for enhanced efficiency.
We first consider the query-contained RR predicate (i.e., atomic condition \textcircled{2}) in Sections 4.1-4.3. Next, we extend our index to support each of the four conditions outlined individually and consider any disjunctive combinations of them in Section 4.4.

\subsection{Multi-Segment Tree Graph Index}
\label{ssec:naive}


Due to the excellent performance of iRangeGraph for RFANN, our initial attempt is to holistically utilize its segment tree-based index. 

First, we construct an iRangeGraph that manages the objects $\{(v_i, r_i) | (v_i, l_i, r_i) \in O \}$, enabling it to handle filter $r_i\in [r_q,+\infty)$.
As to $l_i\in (-\infty,l_q]$, we build multi-segment trees.
For simplicity, we define $O_x = \{o_i | o_i \in O, l_i \le a_x \in A\}$, where $a_x < a_{x+1}$ for each $1 \le x\le |A|$ and $a_{|A|+1} = +\infty$. For each $O_x$, we establish a segment tree based on $r_i$ for each $o_i\in O_x$. Like iRangeGraph, we construct a PG for each node of the segment tree, and denote the graph index of $O_x$ as $G_x$. For simplicity, we denote this new index as multi-segment tree graph (MSTG). 
%
Given a query $q=(v_q,l_q,r_q)$, we first locate $G_x$ where $a_{x} \le l_q < a_{x+1}$ (assuming $a_{|A|+1}=+\infty$). As $G_x$ contains objects in $O_x$, each object $o_i \in O_x$ meeting $l_i\le l_q$ criteria is contained within $G_x$. Subsequently, we employ the segment tree within $G_x$ to identify nodes covering the range $[r_q,+\infty)$. Since the segment tree in $G_x$ is built based on $r_i$ for each $o_i\in O_x$, the identified nodes in $G_x$ adhere to $r_i\ge r_q$ and $l_i\le l_q$ conditions. Given that $l_q \le r_q$ holds for the query, this method effectively captures all nodes satisfying $l_i\le l_q \le r_q\le r_i$.

Therefore, by consolidating the PGs derived from nodes of $G_x$ satisfying $r_i \in [r_q,+\infty)$ into a new PG $\mathcal{G}$, we can execute a $k$-ANNS on $\mathcal{G}$ to retrieve the results of RRANN queries with query-contained filters (atomic condition \textcircled{2}). It is important to note that constructing $\mathcal{G}$ individually for each query is unnecessary, as it can be virtually formed during the search process.
Considering $\mathcal{G}_1,\cdots,\mathcal{G}_p$ as the $p$ PGs derived from $G_x$, we can modify line 5 in Algorithm~\ref{alg:knn_search} to ``\textbf{for} {each $v \in N_{\mathcal{G}}(u)$} \textbf{do}'' for our search on MSTG. Here, $N_{\mathcal{G}}(u)$ is derived from $N_{\mathcal{G}_1}(u)\cup \cdots\cup N_{\mathcal{G}_p}(u)$, and we can guarantee $|N_{\mathcal{G}}(u)| \leq m$ by a pruning strategy that prioritizes neighbors in higher layers close to the tree root.



\stitle{Search Complexity}:
Our MSTG approach guarantees the search efficiency of RRANN with query-contained filters, as supported by Lemma~\ref{lemma:log-segment-tree} shown and proved in Appendix~\ref{sec:proofs}: the number $p$ of involved PGs during the search process ranges up to $O(\log |A|)$.
%
%
%
%
Thus, the sole discrepancy in time complexity between our search algorithm and $k$-ANN search lies at line 5 in Algorithm~\ref{alg:knn_search}. In $k$-ANN search, the complexity at line 5 is $O(md)$, whereas in our method, it extends up to $O(m\log|A| + md)$. Since $\log |A|$ is always much smaller than $d$ in practice, our search has a similar search complexity to the $k$-ANN search on PGs. Moreover, our approach excels in its capability to completely avoid traversing nodes that fail to satisfy the query filter conditions.

\stitle{Index Complexity}:
For the construction of MSTG, we initially sort the objects based on their $l_i$ values. Subsequently, each $O_x$ contains a prefix of the sorted object sequence. For every $O_x$ ($1 \leq x \leq |A|$), we build an index $G_x$, which involves creating a segment tree based on $\{r_i \mid o_i \in O_x \}$ and then constructing a PG on the vectors in each tree node, which needs the same time as iRangeGraph.
Hence, the indexing process requires a time complexity of $O(|A|\cdot T_{iRG})$, where $T_{iRG}$ denotes the construction time of iRangeGraph. Moreover, the space complexity is $O(nm|A|\log |A|)$, where $m$ represents the out-degree limit on each PG and $n$ indicates the number of objects in $O$. Given that $|A|$ may reach up to $n$ in the worst case, both index construction time and size incur significant costs.

\subsection{Merged Multi-Segment Tree Graph Index}
\label{ssec:pstg}

\begin{figure*}[t]
\centering
\includegraphics[width=0.9\linewidth]{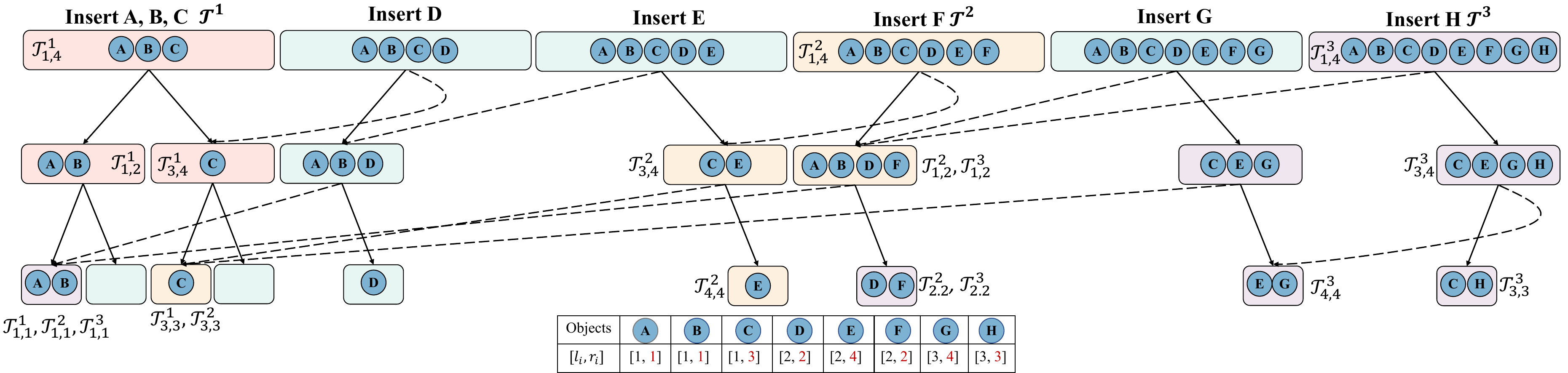}
\vspace{-4mm}
\caption{The illustration of the merged MSTG}
\label{fig:pst}
\vspace{-3mm}
\end{figure*}



In this subsection, we explore the merging of MSTG to reduce its construction costs.
Let us consider the consecutive construction of two graph indices $G_x$ and $G_{x+1}$ in MSTG, which contains objects $\{o_i\in O\mid l_i\le a_x\}$ and $\{o_i\in O\mid l_i\le a_{x+1}\}$ respectively. The disparity in contained objects in $G_x$ and $G_{x+1}$ is $\{o_i\in O\mid l_i=a_{x+1}\}$. Hence, we contemplate creating $G_{x+1}$ by adding the vectors of objects in $\{o_i\in O\mid l_i=a_{x+1}\}$ into $G_x$. For simplicity, we opt to add one vector at a time, allowing for iterative additions.

To add an object $o_i=(v_i,a_{x+1},r_i)$ in $G_x$, we can observe that the majority of nodes within the segment tree of $G_x$ remain unaffected. This is because the segment tree is established based on $\{r_i|o_i=(v_i,l_i,r_i)\in O\}$. Hence, the inclusion of the object's $r_i$ solely impacts $O(\log |A|)$ nodes within the segment tree, i.e., the nodes on the path from the root node to the leaf node containing $r_i$. 

\begin{example}
\label{example-1}
Consider the example depicted in Fig.~\ref{fig:pst}. We insert $D$ into the segment tree $\mathcal{T}^1$, which currently contains three objects $A$, $B$, and $C$. Given that $r_D=2$, the nodes along the path from the root node to the leaf node of $r_D$ represent the ranges $[1,4]$, $[1,2]$, and $[2,2]$ respectively. Hence, we only need to update these three nodes, while the remaining nodes remain unchanged from $\mathcal{T}^1$.
\end{example}

As a result, to construct a new graph index by adding $o_i$ into $G_x$, only $O(\log |A|)$ nodes and their respective PGs need updating. 
The remaining nodes, devoid of redundant storage, can be directly incorporated into the new index by preserving the parent-child relationships between the new graph index and $G_x$.
For example, when inserting $D$ into $\mathcal{T}^1$ as illustrated in Example~\ref{example-1}, the root node representing $[1,4]$ and its left child $[1,2]$ become new nodes due to the insertion of $D$, while the subtree rooted at the node representing $[3,4]$ remains unchanged. Thus, we can directly use a pointer to designate the node $\mathcal{T}^1_{3,4}$ as its right child.
Similarly, for the node representing $[1,2]$,  we designate $\mathcal{T}^1_{1,1}$ as its left child and create a node to serve as its right child.
Thus, by sequentially adding each object of $\{o_i\in O\mid l_i=a_{x+1}\}$ into the current MSTG $G_x$, 
we build the MSTG $G_{x+1}$. More details are presented below.



\begin{algorithm}[t]
\small  
    \SetVline 
    \SetFuncSty{textsf}
    \SetArgSty{textsf}
\caption{\texttt{InsertMSTG}($o_i,\mathcal{T}^x,A,m$)}
\label{alg:build_insert}
\Input{$o_i$: an object, $\mathcal{T}^x$: current segment tree, $A$: the attribute set, $m$: the out-degree limit}
\Output{the new segment tree $\mathcal{T}$}
\State{$l\leftarrow 1$; $r\leftarrow |A|$}
\While{$l< r$}
{
    \State{$\mathcal{T}_{l,r}\leftarrow \mathcal{T}^x_{l,r}\cup \{o_i\}$ and $\mathcal{T}_{l,r}.G\leftarrow$ a PG of objects in $\mathcal{T}_{l,r}$ with out-degree limit $m$}
    \State{\textbf{if} $l=r$ \textbf{then} break}
    \State{$\alpha= \left \lfloor (l+r)/2\right \rfloor$}
    \State{\textbf{if} $r_i\le \alpha$ \textbf{then} $\mathcal{T}_{\alpha+1,r}\leftarrow \mathcal{T}^x_{\alpha+1,r}$; $r\leftarrow \alpha$}
    \State{\textbf{else} $\mathcal{T}_{l,\alpha}\leftarrow \mathcal{T}^x_{l,\alpha}$; $l\leftarrow \alpha+1$}
}
\Return{$\mathcal{T}$}
\end{algorithm}

\stitle{Construction Algorithm}:
It starts from an empty segment tree $\mathcal{T}^0$, where each node in $\mathcal{T}^0$ has already determined its range based on $R=\{r_i | o_i \in O\}$, although it contains no objects, as depicted in Fig.~\ref{fig:pst}. Next, we insert each $o_i \in O$ into MSTG in ascending order of its $l_i$ value, employing its $r_i$ value as the key for segment tree insertion. This process constructs indexes $\mathcal{T}^1, \cdots,\mathcal{T}^{|A|}$, where $\mathcal{T}^x$ corresponds to the graph index $G_{x}$ and manages $O_x$.

When inserting an object $o_i=(v_i,l_i,r_i)$ into the current segment tree $\mathcal{T}^x$, we represent each tree node in $\mathcal{T}^x$ by $\mathcal{T}^x_{l,r}$, where $[l,r]$ indicates its range of $r_i$ values, and denote its PG as $\mathcal{T}^x_{l,r}.G$. For instance, $\mathcal{T}^x_{1,|A|}$ denotes the root node of the segment tree $\mathcal{T}_x$. For a node $\mathcal{T}^x_{l,r}$ (where $l \neq r$ implies a non-leaf node) in $\mathcal{T}^x$, its left-child node is $\mathcal{T}^x_{l,\alpha}$, and its right-child node is $\mathcal{T}^x_{\alpha+1,r}$, where $\alpha= \left \lfloor (l+r)/2\right \rfloor$.
As in Algorithm~\ref{alg:build_insert}, the recursive insertions start at root node $\mathcal{T}^x_{1,|A|}$ (line 1). It reconstructs a new PG for nodes containing a newly added object (line 3), and proceeds to recursively add the object to the left-child or right-child node based on $r_i$ (lines 6-7). The path from its non-recursive child directly points to the segment tree $\mathcal{T}_x$ to streamline construction time.
Next, as shown in Algorithm~\ref{alg:build_ptsg}, we iteratively add objects to the current index. We first create a new empty segment tree $\mathcal{T}^0$ based on $A$ (line 1). We then traverse the objects in the set $O$ in ascending order of $l_i$ (line 3), adding each object to the current segment tree index $\mathcal{T}$ (line 4). Upon encountering $l_i\neq l_{i+1}$ (line 5), the set $\{o_j\mid j\le i\}$ forms a set $O_x$, indicating that the current $\mathcal{T}$ corresponds to $\mathcal{T}^x$ where $a_x=l_i$ (lines 6-7).

\begin{algorithm}[t]
\small  
    \SetVline 
    \SetFuncSty{textsf}
    \SetArgSty{textsf}
\caption{\texttt{ConstructMSTG}($O,A,m$)}
\label{alg:build_ptsg}
\Input{$O$: the object set, $A=\{a_1,\cdots,a_{|A|}\}$: {the attribute set ($a_1< \cdots,a_{|A|}$)}, and $m$: the out-degree limit}
\Output{The merged MSTG $\mathcal{T}^1,\cdots,\mathcal{T}^{|A|}$}
\State{build a segment tree $\mathcal{T}^0$ based on $A$ without objects}
\State{$x\leftarrow 1$}
\For{each $o_i=(v_i,l_i,r_i)\in O$ in ascending order of $l_i$}
{
    \State{$\mathcal{T}\leftarrow$\kw{InsertMSTG}($o_i,\mathcal{T},A,m$)}
    \If{$l_i\neq l_{i+1}$}
    {
    \State{\textbf{while} $a_x\neq l_i$ \textbf{do} $x\leftarrow x+1$}
        \State{$\mathcal{T}^x\leftarrow \mathcal{T}$; $x\leftarrow x+1$}
    }
}
\Return{$\mathcal{T}^1,\cdots,\mathcal{T}^{|A|}$}
\end{algorithm}

\begin{example}
Referring back to Example~\ref{example-1} depicted in Fig.~\ref{fig:pst}, we have $A=\{1,2,3,4\}$, and solid lines indicate newly created pointers to tree nodes, and dashed lines indicate reused tree nodes.
Given that $l_A=l_B=l_C\neq l_D$, the tree containing $A$, $B$, and $C$ is $\mathcal{T}^1$. 
Next, we continue to insert $E$. As $r_E=4$, new tree nodes representing the ranges $[1,4]$, $[3,4]$, and $[4,4]$ are created, while other pointers refer to tree nodes existing before the insertion.
Upon inserting $F$, where $r_F=2$, new tree nodes representing the ranges $[1,4]$, $[1,2]$, and $[2,2]$ are created. After insertion, finding that $l_F\neq l_G$ (line 5 in Algorithm~\ref{alg:build_ptsg}),  since $a_x=2$, the current tree is $\mathcal{T}^2$ (tree nodes highlighted in yellow, with $\mathcal{T}^2_{1,1}$ in purple).
Continuing with the insertion of objects $G$ and $H$, we eventually have the tree $\mathcal{T}^3$ (tree nodes are colored in purple) after inserting $H$. 
\end{example}


\stitle{Search Method:}
The search process remains consistent with the method in the last part. It first locates the index $\mathcal{T}^x$ where $a_x\le l_q < a_{x+1}$ (assuming $a_{|A|+1}=+\infty$), followed by identifying nodes with $r_i \in [r_q,+\infty)$ in the segment tree within $\mathcal{T}^x$. Next, it consolidates PGs in these nodes into $\mathcal{G}$ under the out-degree limit $m$, and executes a $k$-ANNS (Algorithm~\ref{alg:knn_search}) on $\mathcal{G}$ to answer the query.


\stitle{Index Complexity}:
First, let us analyze the building cost of a merged MSTG. Notably, building $\mathcal{T}_{l,r}.G$ in line 3 of Algorithm~\ref{alg:build_insert} does not necessitate building from scratch. It inserts $o_i$ into $\mathcal{T}^x_{l,r}.G$, e.g., requiring only a $k$-ANN search for $v_i$ and a subsequent pruning when the PG is HNSW~\cite{Hnsw}. As each $o_i\in O$ is added into $O(\log |A|)$ nodes, the construction time amounts to $O(n \log |A| T_{insert})$, where $T_{insert}$ denotes the cost of inserting a vector into the PG. When using HNSW, its building cost is the same as iRangeGraph~\cite{iRange}.

Second, for space complexity, upon the insertion of a new object, $O(\log |A|)$ new tree nodes are added, resulting in a total of $O(n\log |A|)$ nodes. In the worst case, all nodes in the current index are contained in the newly added $O(\log |A|)$ tree nodes during each insertion. Hence, the merged MSTG requires a maximum of $O(n^2m \log |A|)$ space, which is much larger than iRangeGraph.
Even if we assume uniform attribute distribution for objects $o_i\in O$, the existing nodes are expected to exist in the newly added $O(1+\frac{1}{2}+\frac{1}{4}+\cdots+\frac{1}{x})=O(1)$ tree nodes on average, where $x$ is the minimal value such that $2^x\ge |A|$. Hence, the space diminishes to $O(nm |A|)$, but it remains much larger than iRangeGraph.


\subsection{Labeled Multi-Segment Tree Graph Index}
\label{ssec:efficient_pstg}

In this part, we delve into compressing our MSTG index. As in Section~\ref{ssec:pstg}, our approach theoretically matches the search performance of iRangeGraph, but suffers from substantial index size. Therefore, our goal here is to achieve a comparable index size to iRangeGraph in theory, while not losing index information during compression to ensure the search performance.

In Section~\ref{ssec:pstg}, we merge identical tree nodes, but similar tree nodes that continue to be observable in MSTG.
For example, in Fig.~\ref{fig:pst}, the tree node representing the range $[3,4]$ differs only in $H$ before and after inserting $H$.
The primary reason for this lies in line 3 of Algorithm~\ref{alg:build_insert}, where the PG constructed for the tree node $\mathcal{T}_{l,r}$ is based on nodes $\mathcal{T}^x_{l,r}\cup\{o_i\}$, with the sole distinction being added object $o_i$. Consequently, the PG $\mathcal{T}_{l,r}.G$ closely resembles $\mathcal{T}^x_{l,r}.G$. For instance, when utilizing HNSW~\cite{Hnsw} as the PG in index, $\mathcal{T}_{l,r}.G$ incorporates an additional node with $m$ edges into $\mathcal{T}^x_{l,r}.G$. 
The disparities between $\mathcal{T}_{l,r}.G$ and $\mathcal{T}^x_{l,r}.G$ amount to one node and at most $m$ edges. Therefore, our approach only stores these differences without the need to fully restore an entire graph index.

\begin{algorithm}[t]
\small  
    \SetVline 
    \SetFuncSty{textsf}
    \SetArgSty{textsf}
\caption{\texttt{InsertLabelHNSW}($G,o_i,m,ef_{con}$)}
\label{alg:build_pg}
\Input{$G$: current HNSW, an object $o_i=(v_i,l_i,r_i)$, $m$: the out-degree limit, and a parameter $ef_{con}$ for index}
\Output{The new HNSW index ${G}'$}
\State{$\mathcal{C}\leftarrow$\kw{KANNSearch}($G,v_i,ef_{con},ef_{con},ep$)}
\State{utilize RNG pruning strategy to ensure $|\mathcal{C}|\le m$}
\State{$G'\leftarrow G\cup \{v_i\}$; $x\leftarrow j$ s.t. $l_i=a_j\in A$}
\For{each $u\in \mathcal{C}$}
{
    \State{add edges $(u,v_i)$ and $(v_i,u)$ into $G'$ with label $(x,+\infty)$}
    \If{$|N_{G'}(u)|> m$}
    {
        \State{$W \leftarrow$ the neighbors pruned by RNG pruning on $N_{G'}(u)$}
        \For{each $w \in W$} 
        {
            \State{$(b,e)\leftarrow$ the label of edge $(u,w)$}
            \State{set the label of edge $(u,w)$ as $(b,x-1)$ in $G'$}
        }
    }
}
\State{\Return{$G'$}}
\end{algorithm}

To be specific, the tree node represents a range $[l,r]$ in the segment tree from $\mathcal{T}^0$ to $\mathcal{T}^{|A|}$, involving the incremental insertion of objects in ascending order of $l_i$. For instance, the PG from $\mathcal{T}^x_{l,r}$ to $\mathcal{T}^{x+1}_{l,r}$ involves adding objects $\{o_i=(v_i,l_i,r_i)\in O\mid l_i=a_{x+1}\}$. Hence, we opt to utilize HNSW as the PG in our MSTG due to its inherent support for vector insertions.

As shown in Algorithm~\ref{alg:build_pg}, the addition of a single vector to the existing HNSW $G$ comprises three steps~\cite{Hnsw}: (1) executing an $ef_{con}$-ANNS on $G$ for the newly inserted vector $v_i$, where $ef_{con}$ is a construction parameter of HNSW (line 1); (2) employing the RNG pruning strategy~\cite{FastPG} to reduce $ef_{con}$-ANN to at most $m$ ones (line 2), denoted as $\mathcal{C}$, and inserting $|\mathcal{C}|$ edges from $v_i$ to nodes in $\mathcal{C}$ in $G'$ (lines 3-5); (3) linking the edges from nodes in $N_{G'}(v_i)$ to the node $v_i$ and subsequently applying the RNG pruning strategy to keep these nodes at most $m$ out-neighbors if necessary (lines 5-9).
Thus, an edge inserted into $G'$ (line 5) might be removed later due to the RNG pruning strategy (line 7). Instead of maintaining $G$ and $G'$ separately, we opt to attach a label $(b,e)$ to each edge. This label signifies that the edge only exists in $\mathcal{T}^b,\mathcal{T}^{b+1},\cdots,\mathcal{T}^e$, where we identify a value $x$ such that $l_i=a_x$ (line 3) and assign the corresponding edges a label of $x$ (lines 5,10).
Therefore, storing $G$ becomes unnecessary when we have $G'$. Likewise, in Algorithm~\ref{alg:build_ptsg}, we can omit $\mathcal{T}^1,\cdots,\mathcal{T}^{|A|-1}$ since $\mathcal{T}^{|A|}$ contains all edges included in $\mathcal{T}^1,\cdots,\mathcal{T}^{|A|-1}$, and the labels of the edges can distinguish which multi-segment tree index they belong to.
We can prove that our labeled MSTG retains all information in MSTG, the details are shown in Theorem~\ref{theo:mstg-same}, which is included in Appendix~\ref{sec:proofs}.



{

\stitle{Index Complexity}:
For the construction time complexity, we do not introduce any additional operations but solely add labels when adding edges. Hence, the time cost remains consistent with the merged MSTG, aligning with iRangeGraph.
For the space complexity, adding a new object into the existing index results in the addition of $O(\log |A|)$ nodes in MSTG. Within the PG of each node, as previously discussed, differences may arise on a maximum of $O(m)$ edges. Consequently, the overall index size amounts to $O(nm\log |A|)$, mirroring that of iRangeGraph.
{Thus, in theory, we achieve equivalent index size and construction time to iRangeGraph, while completely avoiding the traversal of nodes that do not satisfy the query filter.}


\stitle{Discussion on Using Other Proximity Graphs}:
It is feasible to integrate other SOTA PG approaches into our MSTG, e.g., NSG~\cite{Nsg}, $\tau$-MNG~\cite{taumg}, CSPG~\cite{YangCZ24}, and ALMG~\cite{ALMG}. The differentiating factor is that these graphs do not inherently facilitate vector insertions. However, numerous approaches exist to support PG maintenance~\cite{abs-2105-09613,milvus,abs-2503-00402,XieYL25}. Thus, by leveraging these techniques, any PG can be integrated into our MSTG index.




\subsection{RRANN with Arbitrary RR Predicates}
\label{ssec:one-rrann}

Here, we explore ways to handle the remaining three conditions by slight modifications to MSTG.

\stitle{\textcircled{1} Query Left-Overlap}: An object $o_i=(v_i,l_i,r_i)$ satisfies the predicate $l_i\leq l_q\leq r_i\leq r_q$, which could be divided into two parts: $l_i\leq l_q$ and $l_q\leq r_i\leq r_q$. Therefore, we can directly utilize the MSTG built for case~\textcircled{2}. We first determine $x$ s.t. $a_x\leq l_q < a_{x+1}$ (assuming $a_{|A|+1}=+\infty$), ensuring $\mathcal{T}^x$ contains all the objects satisfying $l_i\leq l_q$. Notably, the segment tree within $\mathcal{T}^x$ is built on $r_i$, allowing the condition $l_q\leq r_i\leq r_q$ to be covered by executing a range query $[l_q,r_q]$ on the segment tree. According to Lemma~\ref{lemma:log-segment-tree}, a maximum of $O(\log n)$ nodes in the segment tree can cover the range $[l_q,r_q]$. Hence, the remaining steps align with those for case \textcircled{2}, involving a $k$-ANN search on the PG virtually built from the PGs w.r.t. the segment tree nodes intersecting with $[l_q,r_q]$.


\stitle{\textcircled{3} Query Right-Overlap}:
In this case, an object $o_i$ satisfies the query predicate when $l_q\leq l_i\leq r_q\leq r_i$, which contains two parts: $l_q\leq l_i\leq r_q$ and $r_q\leq r_i$. 
We solve it by sequentially constructing $\mathcal{T}'^{|A|},\mathcal{T}'^{|A|-1},\ldots,\mathcal{T}'^{1}$ through the insertion of objects $o_i$ in descending order of $r_i$, where $\mathcal{T}'^x$ comprises objects $\{o_i \mid r_i\geq a_x\}$. Each object $o_i$ is then inserted into $\mathcal{T}'^x$ if $r_i=a_x$, subsequently being inserted into $O(\log |A|)$ PGs on the segment tree nodes whose range encompasses $l_i$. 
For a query $q=(v_q,l_q,r_q)$, the search first locates $x$ s.t. $a_{x-1}< r_q\leq a_x$ (assuming $a_0=-\infty$), whereby $\mathcal{T}'^x$ comprises all objects in $\{o_i \mid r_i\geq r_q\}$. Then, it finds the segment tree nodes intersecting $[l_q,r_q]$ in $\mathcal{T}'^x$, followed by a $k$-ANN search on the virtually merged PG of these nodes.

\stitle{\textcircled{4} Query-Containing}:
In this case, an object $o_i$ satisfies the query predicate $l_q\leq l_i\leq r_i\leq r_q$.
Since $l_i\leq r_i$ always holds, we can equivalently rephrase the condition as $l_q\leq l_i$ and $r_i\leq r_q$. Hence, we can sequentially construct $\mathcal{T}''^{|A|},\mathcal{T}''^{|A|-1},\ldots,\mathcal{T}''^{1}$ by inserting objects $o_i$ in descending order of $l_i$, where $\mathcal{T}''^x$ includes objects $\{o_i=(v_i,l_i,r_i)\in O\mid l_i\geq a_x\}$. Then, similar to $\mathcal{T}^x$, each $o_i$ is inserted into $O(\log |A|)$ nodes' PGs of $\mathcal{T}''^x$ based on $r_i$. 
For a query $q=(v_q,l_q,r_q)$, the process first determines $x$ s.t. $a_{x-1}< l_q\leq a_x$ (assuming $a_0=-\infty$), whereby $\mathcal{T}''^x$ contains all objects in $\{o_i=(v_i,l_i,r_i)\in O\mid l_i\geq l_q\}$. Then, it identifies the segment tree nodes intersecting with the range $(-\infty,r_q]$ in $\mathcal{T}''^x$, followed by a $k$-ANNS on the virtually merged PG of these nodes.





As discussed above, addressing RRANN with 4 atomic RR predicates requires 3 variants of MSTG indexes. A direct approach to a combined RR predicate, i.e., a disjunctive combination of these filters, builds three MSTG indexes separately, addresses each atomic filter individually, and finally merges the $k$-ANN results from each atomic filter to derive the final outcomes. However, this method is inefficient for two primary reasons: (1) the construction of three MSTG indexes incurs a substantial index cost, and (2) it requires multiple individual queries for a single RRANN query. For example, a RR predicate  \textcircled{1}$\vee$\textcircled{2}$\vee$\textcircled{3}$\vee$\textcircled{4} demands four RRANN queries w.r.t. four atomic filters.
The following theorem states that only a maximum of two distinct MSTG indexes are enough for processing RRANN queries with combined RR predicates; hence, we can efficiently address RRANN queries with arbitrary predicates by at most two searches. The proof is included in Appendix~\ref{sec:proofs}.



\begin{theorem}
For an RRANN query involving any combined RR predicates, a maximum of two MSTG indexes and no more than two distinct searches are necessary to be conducted.
\label{theo:combine-predicate}
\end{theorem}

\begin{figure*}[t]
  \centering
  \includegraphics[width=0.5\linewidth]{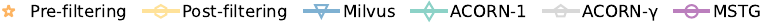}

  \includegraphics[width=\linewidth]{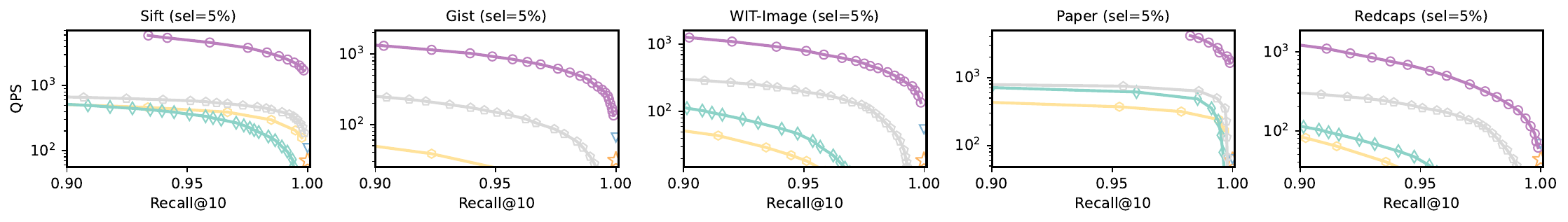}
    \includegraphics[width=\linewidth]{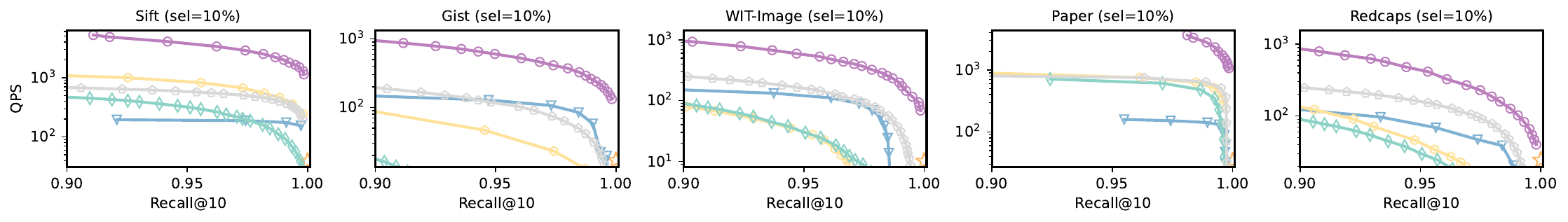}
  \vspace{-8mm}
  \caption{Overall query performance of RRANN (Exp. 1)}
  \label{fig:overall_interval}
  \vspace{-3mm}
\end{figure*}




\section{Experiments}
\label{sec:exp}

In this section, we conduct extensive experiments on real-world datasets and report our findings.



\begin{figure}[t]
  \centering
    \includegraphics[width=\linewidth]{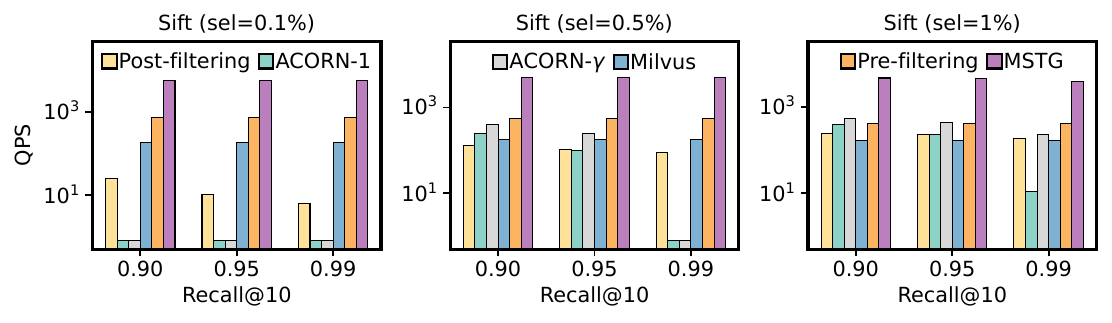}
  \vspace{-8mm}
  \caption{RRANN performance with low selectivities (Exp. 1)}
  \label{fig:overall_interval_low_sel}
  \vspace{-3mm}
\end{figure}

\begin{figure}[t]
  \centering
\includegraphics[width=0.85\linewidth]{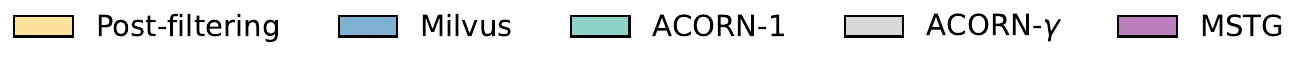}
  
  \includegraphics[width=\linewidth]{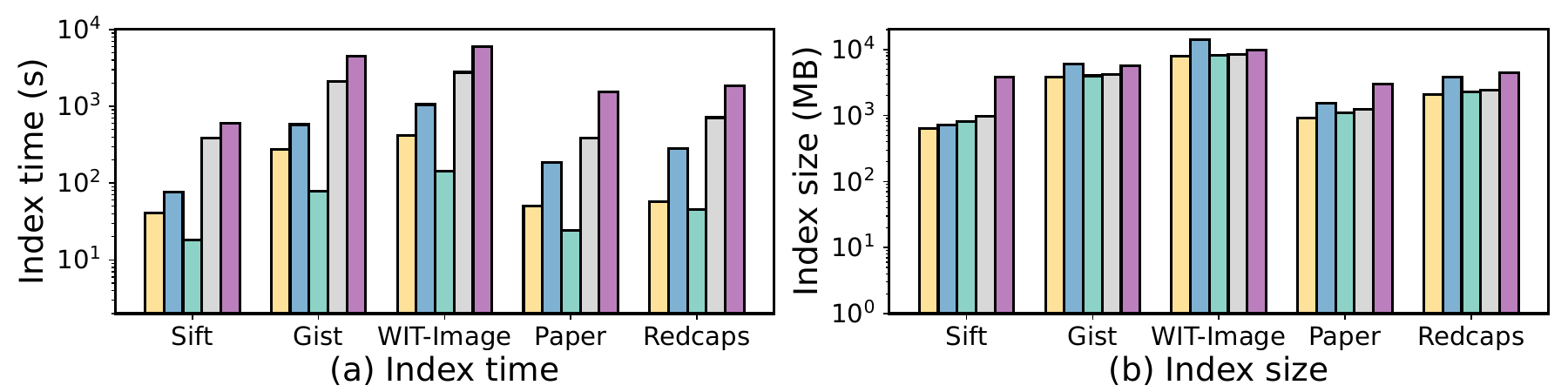}
  \vspace{-5mm}
  \caption{Indexing costs of RRANN queries (Exp. 2)}
  \label{fig:interval_time_and_size}
  \vspace{-3mm}
\end{figure}

\begin{figure*}[t]
  \centering
  \includegraphics[width=0.85\linewidth]{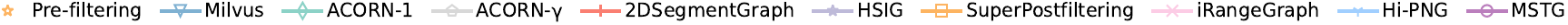}


  \includegraphics[width=\linewidth]{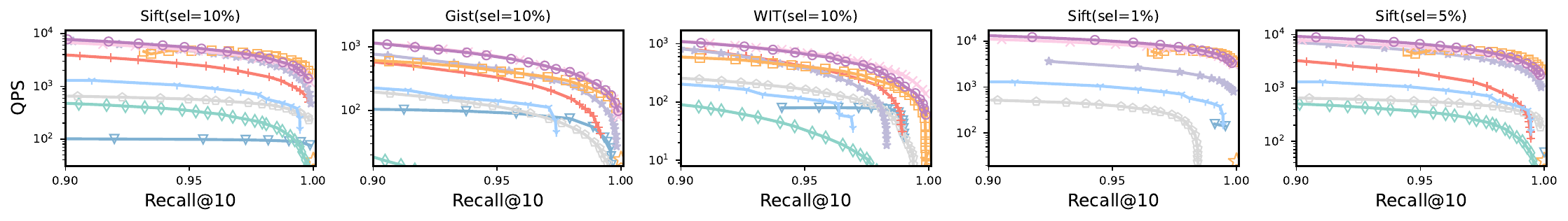}
  
  \vspace{-5mm}
  \caption{Query performance on RFANN queries (Exp. 3)}
  \label{fig:overall_range}
  \vspace{-3mm}
\end{figure*}

\noindent \textbf{Datasets}:
We utilize six real-world datasets in various domains, including image (Sift~\cite{sift}, Gist~\cite{sift}, WIT-Image~\cite{WIT}), text (Paper~\cite{NHQ}), and image-text multimodality (Redcaps~\cite{Redcaps}). 
The statistics of the datasets and their queries are included in Appendix~\ref{sec:exp_setting}.

\noindent \textbf{Compared Algorithms and Parameters}:
We first compare our approach {MSTG}, with RRANN methods, as discussed in Section~\ref{sec:limit}. The compared methods include {(1) ACORN~\cite{ACORN}}, {(2) Post-filtering~\cite{ADBV,milvus}}, {(3) Pre-filtering~\cite{ADBV,milvus,vbase}}, and {(4) Milvus~\cite{milvus}}. 
Since RFANN could be treated as a special case of our RRANN problem, our method naturally supports RFANN queries and thus we compare MSTG with the SOTA RFANN methods: {(5) 2DSegmentGraph~\cite{SeRF}}, {(6) {HSIG}~\cite{UNIFY}}, {(7) {SuperPostfiltering}~\cite{WST}}, and {(8) {iRangeGraph}~\cite{iRange}}.
As TSANN and IFANN are also two special cases of RRANN, we compare MSTG with {(9) TS-Graph}~\cite{TSANN} for TSANN queries and {(10) Hi-PNG}~\cite{Hi-PNG} for IFANN queries, respectively. 
Their parameter settings and details can be found in Appendix~\ref{sec:exp_setting}.

\noindent \textbf{Performance Indicators}:
\sloppy
We employ recall at $k$ (Recall@k) and relative distance error (RDE) to measure the search accuracy. Recall@k is the ratio of successfully retrieved ground truth $k$-NN to $k$-ANN, while RDE is the relative distance error between ground truth $k$-NN and $k$-ANN. { For a query $q$, RDE is computed as $1/k\sum_{i=1}^k (\delta(q,p_i)/\delta(q,p_i^*))-1$, where $p_i$ is the $i$-th neighbor in retrieved $k$-ANN and $p_i^*$ is the $i$-th ground truth neighbor. }
Search efficiency is assessed by the number of queries processed per second (QPS). 
All experiments are averaged over five independent runs.

The environment of experiments are shown in Appendix~\ref{sec:exp_setting}.



\stitle{Exp. 1: Query Performance of RRANN Queries.} 
We evaluate the query performance of our method and other baselines on RRANN queries, where the RR predicates is set as \textcircled{1}$\vee$\textcircled{2}$\vee$\textcircled{3}$\vee$\textcircled{4}.
Fig.~\ref{fig:overall_interval} presents the QPS–recall curves with 5\% and 10\% selectivities. Our method MSTG, consistently surpasses all baselines, particularly achieving 5.2x–12.5x higher QPS with the same recall compared to the best competitor ACORN-$\gamma$, across all datasets with Recall@10 as 0.99 and selectivity as 5\%.
Moreover, MSTG stands out as the sole method capable of attaining a high QPS at high recall. 
On Gist with a selectivity of 5\%, Post-filtering falls short of achieving a recall of 0.95, and ACORN-1 struggles to reach a recall of 0.9 with the same QPS level as Pre-filtering. 
As selectivity rises to 10\%, these baselines show modest performance improvements due to enhanced graph connectivity, but still face challenges in achieving a high recall level.
{Similarly, MSTG presents significant advantages in the search performance of low-selectivity RRANN queries, as presented in Figure~\ref{fig:overall_interval_low_sel} with $sel$ set as 0.1\%, 0.5\% and 1\%, respectively.}
{Moreover, our method presents similar advantages over the baselines, as measured by RDE, as shown in Fig.~\ref{fig:overall_interval_rde} of Appendix~\ref{sec:exp_others}. }

  



\begin{figure}[!t]
  \centering
\includegraphics[width=\linewidth]{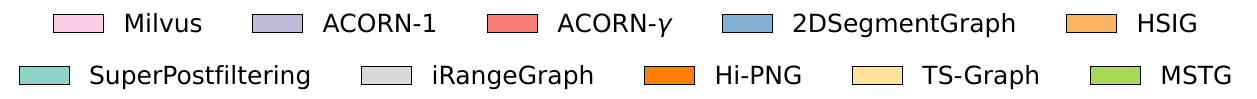}
\includegraphics[width=\linewidth]{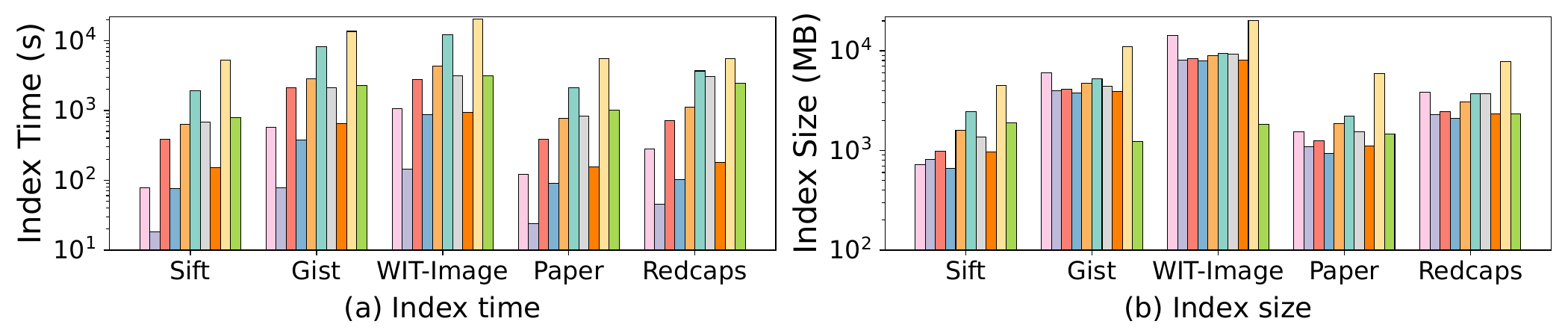}
  \vspace{-8mm}
  \caption{Indexing costs (Exp. 3\&4\&5)}
  \label{fig:range_time_and_size}
  \vspace{-3mm}
\end{figure}

\stitle{Exp. 2: Indexing Cost.}
We evaluate the indexing cost of all methods on RRANN queries in terms of construction time and index size, as shown in Fig.~\ref{fig:interval_time_and_size}.
General-purpose approaches such as Post-filtering, Milvus, ACORN-1, and ACORN-$\gamma$ exhibit relatively low construction overhead, but leads to poor RRANN search performance. Although Milvus uses the same HNSW parameters as Post-filtering, its construction time is higher due to the additional scalar index. Compared to ACORN-$\gamma$, MSTG trades slightly more indexing time for notably better query performance. The index size exhibits a similar trend, i.e., most general-purpose methods occupy relatively little space. However, on Gist and WIT-Image, Milvus consumes more index space due to its segmented storage.
Overall, although MSTG does not excel in indexing time or index size, its superior query performance makes it a worthwhile trade-off.

  


\stitle{Exp. 3: Performance of RFANN Queries.}
Since RFANN is a special case of RRANN, MSTG naturally supports RFANN queries. Hence, we compare it with the SOTA RFANN methods in Fig.~\ref{fig:overall_range} with the query selectivity as 10\%. In general, MSTG significantly outperforms all the baselines except iRangeGraph in RFANN search performance. MSTG and iRangeGraph are theoretically expected to yield comparable performance, which is verified in this experiment. Both MSTG and iRangeGraph build the PG for the subset satisfying the query filter in an online and virtual manner and conduct $k$-ANNS on the PG without verifying out-of-range candidates, which leads to their superior performance. 
Moreover, we report the indexing time and index size in Fig.~\ref{fig:range_time_and_size}. The indexing cost of MSTG is comparable to iRangeGraph, but higher than others, except SuperPostfiltering. The index size of MSTG is slightly larger than iRangeGraph, since MSTG contains extra label information.

\stitle{Exp. 4: Performance of IFANN Queries.}
Since IFANN is a special case of RRANN, MSTG naturally supports IFANN queries. Hence, we compare it with the SOTA IFANN method, i.e., Hi-PNG~\cite{Hi-PNG}, in Fig.~\ref{fig:ifann}. MSTG significantly outperforms Hi-PNG. This is because Hi-PNG has to conduct $k$-ANNS on multiple PGs, rather than only one in MSTG, and verify candidates that do not satisfy the query filter.
We compare the index costs in index time and index size bewteen our method and Hi-PNG in Figure~\ref{fig:range_time_and_size}.
Compared with Hi-PNG, although more time and space are required to build our index, as objects may need to be stored in multiple PGs, this process guarantees significantly improved performance.

\stitle{Exp. 5: Performance of TSANN Queries.}
Since TSANN is a special case of RRANN, MSTG naturally supports TSANN queries. Here, we compare MSTG with the SOTA TSANN method, i.e., TS-Graph in Fig.~\ref{fig:tsann}. We can see that MSTG significantly outperforms TS-Graph.
Moreover, our index building time and space requirements are significantly lower compared to TS-Graph, as shown in Figure~\ref{fig:range_time_and_size}. To be specific, on Gist, TS-Graph needs over 10,000 seconds to construct its index of size 11.44 GB, whereas we complete the process in 2,300 seconds with an index of size 1.21 GB.

\begin{figure}[t]
  \centering
  \includegraphics[width=\linewidth]{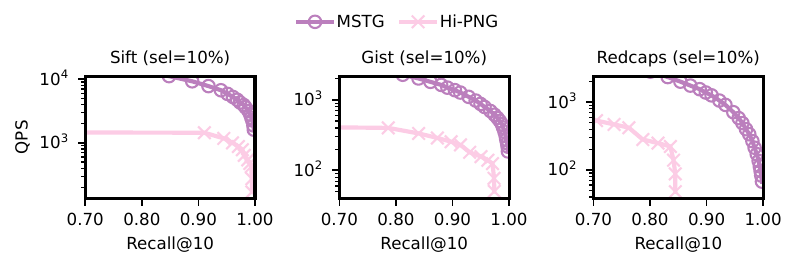}
  \vspace{-8mm}
  \caption{Query performance of IFANN queries (Exp. 4)}
  \label{fig:ifann}
  \vspace{-3mm}
\end{figure}

\begin{figure}[t]
  \centering
  \includegraphics[width=\linewidth]{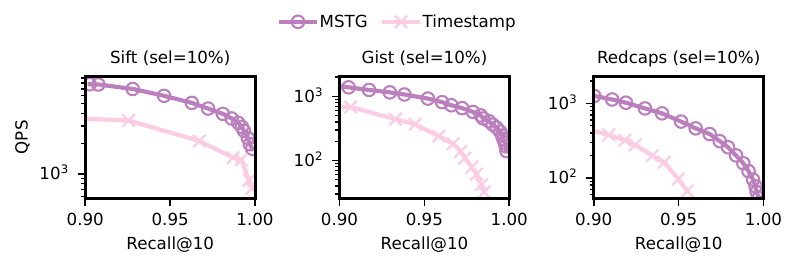}
  \vspace{-8mm}
  \caption{Query performance of TSANN queries (Exp. 5)}
  \label{fig:tsann}
  \vspace{-3mm}
\end{figure}

\begin{figure}[t]
  \centering
  \includegraphics[width=\linewidth]{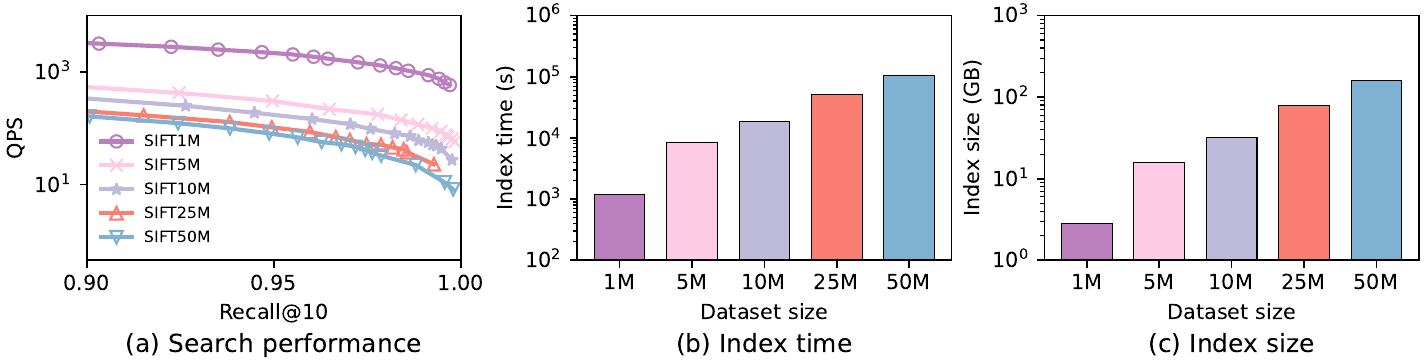}
  \vspace{-5mm}
  \caption{Scalability evaluation of MSTG (Exp. 6)}
  \label{fig:scalability_serach}
  \vspace{-3mm}
\end{figure}

\noindent \textbf{Exp. 6: Scalability of Our Method.}
We test the scalability of our method by sampling various-sized subsets of Sift50M, the large-scale dataset widely used for scalability tests. We show the search performance and index cost in Fig.~\ref{fig:scalability_serach}. We can see that the search performance gradually decreases and the index cost in both time and space steadily increases as the data size grows. Hence, MSTG could be well scaled to large data.

\stitle{Other Experiments:} Due to the space limit, we put other important experiments in {Appendix~\ref{sec:exp_others}}. 
Specifically, we investigate the impact of query selectivity, attribute distribution, and the cardinality of $A$ on the performance of RRANN queries in Exp.s 7, 8, and 10, respectively. We also explore the effects of parameter $k$, $ef_{con}$, and $M$ on the performance of our method in Exp.s 11-13, respectively. We compare our MSTG with the Oracle-HNSW, which is built solely on $O[R_q]$ in Exp. 8, to demonstrate the effectiveness of our method. We also present the comparisons of RRANN methods with relative distance error (RDE) as the accuracy measure. 

\section{Conclusion}
\label{sec:conclu}

{In this paper, we propose MSTG index to solve the RRANN problem, the $k$-ANNS with various RR predicates, which is the first attempt to solve this problem in a general view, to the best of our knowledge.
Extensive experiments demonstrate that our approach significantly outperforms competitors in RRANN search performance by up to 12.5x efficiency with the same recall level. For RFANN, our approach has comparable performance compared to the SOTA method iRangeGraph, while for TSANN and IFANN, our method achieves much superior search performance compared to the SOTA methods by up to more than one order of magnitude on efficiency.}

\begin{acks}
This work was supported in part by the Jing-Jin-Ji Regional Integrated Environmental Improvement-National Science and Technology Major Project of Ministry of Ecology and Environment of China (No. 2025ZD1200600), and the National Natural Science Foundation of China (No. 62372352).
\end{acks}


\balance
\bibliographystyle{ACM-Reference-Format}
\bibliography{main}


\begin{thebibliography}{39}


\ifx \showCODEN    \undefined \def \showCODEN     #1{\unskip}     \fi
\ifx \showDOI      \undefined \def \showDOI       #1{#1}\fi
\ifx \showISBNx    \undefined \def \showISBNx     #1{\unskip}     \fi
\ifx \showISBNxiii \undefined \def \showISBNxiii  #1{\unskip}     \fi
\ifx \showISSN     \undefined \def \showISSN      #1{\unskip}     \fi
\ifx \showLCCN     \undefined \def \showLCCN      #1{\unskip}     \fi
\ifx \shownote     \undefined \def \shownote      #1{#1}          \fi
\ifx \showarticletitle \undefined \def \showarticletitle #1{#1}   \fi
\ifx \showURL      \undefined \def \showURL       {\relax}        \fi
\providecommand\bibfield[2]{#2}
\providecommand\bibinfo[2]{#2}
\providecommand\natexlab[1]{#1}
\providecommand\showeprint[2][]{arXiv:#2}

\bibitem[\protect\citeauthoryear{??}{sif}{2010}]%
        {sift}
 \bibinfo{year}{2010}\natexlab{}.
\newblock \bibinfo{title}{Datasets for approximate nearest neighbor search}.
\newblock \bibinfo{howpublished}{\url{http://corpus-texmex.irisa.fr/}}.
\newblock


\bibitem[\protect\citeauthoryear{Allen}{Allen}{1983}]%
        {Allen83}
\bibfield{author}{\bibinfo{person}{James~F. Allen}.}
  \bibinfo{year}{1983}\natexlab{}.
\newblock \showarticletitle{Maintaining Knowledge about Temporal Intervals}.
\newblock \bibinfo{journal}{\emph{Commun. {ACM}}} \bibinfo{volume}{26},
  \bibinfo{number}{11} (\bibinfo{year}{1983}), \bibinfo{pages}{832--843}.
\newblock


\bibitem[\protect\citeauthoryear{Azizi, Echihabi, and Palpanas}{Azizi
  et~al\mbox{.}}{2023}]%
        {AziziEP23}
\bibfield{author}{\bibinfo{person}{Ilias Azizi}, \bibinfo{person}{Karima
  Echihabi}, {and} \bibinfo{person}{Themis Palpanas}.}
  \bibinfo{year}{2023}\natexlab{}.
\newblock \showarticletitle{Elpis: Graph-Based Similarity Search for Scalable
  Data Science}.
\newblock \bibinfo{journal}{\emph{Proc. {VLDB} Endow.}} \bibinfo{volume}{16},
  \bibinfo{number}{6} (\bibinfo{year}{2023}), \bibinfo{pages}{1548--1559}.
\newblock


\bibitem[\protect\citeauthoryear{Azizi, Echihabi, and Palpanas}{Azizi
  et~al\mbox{.}}{2025}]%
        {AziziEP25}
\bibfield{author}{\bibinfo{person}{Ilias Azizi}, \bibinfo{person}{Karima
  Echihabi}, {and} \bibinfo{person}{Themis Palpanas}.}
  \bibinfo{year}{2025}\natexlab{}.
\newblock \showarticletitle{Graph-Based Vector Search: An Experimental
  Evaluation of the State-of-the-Art}.
\newblock \bibinfo{journal}{\emph{Proc. {ACM} Manag. Data}}
  \bibinfo{volume}{3}, \bibinfo{number}{1} (\bibinfo{year}{2025}),
  \bibinfo{pages}{43:1--43:31}.
\newblock


\bibitem[\protect\citeauthoryear{Cai, Shi, Chen, and Zheng}{Cai
  et~al\mbox{.}}{2024}]%
        {UNG}
\bibfield{author}{\bibinfo{person}{Yuzheng Cai}, \bibinfo{person}{Jiayang Shi},
  \bibinfo{person}{Yizhuo Chen}, {and} \bibinfo{person}{Weiguo Zheng}.}
  \bibinfo{year}{2024}\natexlab{}.
\newblock \showarticletitle{Navigating Labels and Vectors: A Unified Approach
  to Filtered Approximate Nearest Neighbor Search}.
\newblock \bibinfo{journal}{\emph{Proceedings of the ACM on Management of
  Data}} \bibinfo{volume}{2}, \bibinfo{number}{6} (\bibinfo{year}{2024}),
  \bibinfo{pages}{1--27}.
\newblock


\bibitem[\protect\citeauthoryear{Deng, You, Xiang, Li, Yuan, Hong, Zheng, Li,
  Li, Liu, Mouratidis, Yiu, Li, Shen, Mao, and Tang}{Deng
  et~al\mbox{.}}{2025}]%
        {abs-2504-10326}
\bibfield{author}{\bibinfo{person}{Yangshen Deng}, \bibinfo{person}{Zhengxin
  You}, \bibinfo{person}{Long Xiang}, \bibinfo{person}{Qilong Li},
  \bibinfo{person}{Peiqi Yuan}, \bibinfo{person}{Zhaoyang Hong},
  \bibinfo{person}{Yitao Zheng}, \bibinfo{person}{Wanting Li},
  \bibinfo{person}{Runzhong Li}, \bibinfo{person}{Haotian Liu},
  \bibinfo{person}{Kyriakos Mouratidis}, \bibinfo{person}{Man~Lung Yiu},
  \bibinfo{person}{Huan Li}, \bibinfo{person}{Qiaomu Shen},
  \bibinfo{person}{Rui Mao}, {and} \bibinfo{person}{Bo Tang}.}
  \bibinfo{year}{2025}\natexlab{}.
\newblock \showarticletitle{AlayaDB: The Data Foundation for Efficient and
  Effective Long-context {LLM} Inference}.
\newblock \bibinfo{journal}{\emph{CoRR}}  \bibinfo{volume}{abs/2504.10326}
  (\bibinfo{year}{2025}).
\newblock


\bibitem[\protect\citeauthoryear{Desai, Kaul, Aysola, and Johnson}{Desai
  et~al\mbox{.}}{2021}]%
        {Redcaps}
\bibfield{author}{\bibinfo{person}{Karan Desai}, \bibinfo{person}{Gaurav Kaul},
  \bibinfo{person}{Zubin Aysola}, {and} \bibinfo{person}{Justin Johnson}.}
  \bibinfo{year}{2021}\natexlab{}.
\newblock \showarticletitle{Redcaps: Web-curated image-text data created by the
  people, for the people}.
\newblock \bibinfo{journal}{\emph{arXiv preprint arXiv:2111.11431}}
  (\bibinfo{year}{2021}).
\newblock


\bibitem[\protect\citeauthoryear{Engels, Landrum, Yu, Dhulipala, and
  Shun}{Engels et~al\mbox{.}}{2024}]%
        {WST}
\bibfield{author}{\bibinfo{person}{Josh Engels}, \bibinfo{person}{Ben Landrum},
  \bibinfo{person}{Shangdi Yu}, \bibinfo{person}{Laxman Dhulipala}, {and}
  \bibinfo{person}{Julian Shun}.} \bibinfo{year}{2024}\natexlab{}.
\newblock \showarticletitle{Approximate Nearest Neighbor Search with Window
  Filters}. In \bibinfo{booktitle}{\emph{ICML}}. \bibinfo{pages}{12469 --
  12490}.
\newblock


\bibitem[\protect\citeauthoryear{Fu, Xiang, Wang, and Cai}{Fu
  et~al\mbox{.}}{2019}]%
        {Nsg}
\bibfield{author}{\bibinfo{person}{Cong Fu}, \bibinfo{person}{Chao Xiang},
  \bibinfo{person}{Changxu Wang}, {and} \bibinfo{person}{Deng Cai}.}
  \bibinfo{year}{2019}\natexlab{}.
\newblock \showarticletitle{Fast approximate nearest neighbor search with the
  navigating spreading-out graph}.
\newblock \bibinfo{journal}{\emph{PVLDB}} \bibinfo{volume}{12},
  \bibinfo{number}{5} (\bibinfo{year}{2019}), \bibinfo{pages}{461--474}.
\newblock


\bibitem[\protect\citeauthoryear{Gollapudi, Karia, Sivashankar, Krishnaswamy,
  Begwani, Raz, Lin, Zhang, Mahapatro, Srinivasan, et~al\mbox{.}}{Gollapudi
  et~al\mbox{.}}{2023}]%
        {Filtered-diskann}
\bibfield{author}{\bibinfo{person}{Siddharth Gollapudi}, \bibinfo{person}{Neel
  Karia}, \bibinfo{person}{Varun Sivashankar}, \bibinfo{person}{Ravishankar
  Krishnaswamy}, \bibinfo{person}{Nikit Begwani}, \bibinfo{person}{Swapnil
  Raz}, \bibinfo{person}{Yiyong Lin}, \bibinfo{person}{Yin Zhang},
  \bibinfo{person}{Neelam Mahapatro}, \bibinfo{person}{Premkumar Srinivasan},
  {et~al\mbox{.}}} \bibinfo{year}{2023}\natexlab{}.
\newblock \showarticletitle{Filtered-diskann: Graph algorithms for approximate
  nearest neighbor search with filters}. In
  \bibinfo{booktitle}{\emph{Proceedings of the ACM Web Conference 2023}}.
  \bibinfo{pages}{3406--3416}.
\newblock


\bibitem[\protect\citeauthoryear{Jiang, Yang, Zhang, Hou, Shi, Zhou, Li, and
  Wang}{Jiang et~al\mbox{.}}{2025}]%
        {JiangYZHSZLW25}
\bibfield{author}{\bibinfo{person}{Mengxu Jiang}, \bibinfo{person}{Zhi Yang},
  \bibinfo{person}{Fangyuan Zhang}, \bibinfo{person}{Guanhao Hou},
  \bibinfo{person}{Jieming Shi}, \bibinfo{person}{Wenchao Zhou},
  \bibinfo{person}{Feifei Li}, {and} \bibinfo{person}{Sibo Wang}.}
  \bibinfo{year}{2025}\natexlab{}.
\newblock \showarticletitle{{DIGRA:} {A} Dynamic Graph Indexing for Approximate
  Nearest Neighbor Search with Range Filter}.
\newblock \bibinfo{journal}{\emph{Proc. {ACM} Manag. Data}}
  \bibinfo{volume}{3}, \bibinfo{number}{3} (\bibinfo{year}{2025}),
  \bibinfo{pages}{148:1--148:26}.
\newblock


\bibitem[\protect\citeauthoryear{Li, Zhang, Sun, Wang, Li, Zhang, and Lin}{Li
  et~al\mbox{.}}{2019}]%
        {Dpg}
\bibfield{author}{\bibinfo{person}{Wen Li}, \bibinfo{person}{Ying Zhang},
  \bibinfo{person}{Yifang Sun}, \bibinfo{person}{Wei Wang},
  \bibinfo{person}{Mingjie Li}, \bibinfo{person}{Wenjie Zhang}, {and}
  \bibinfo{person}{Xuemin Lin}.} \bibinfo{year}{2019}\natexlab{}.
\newblock \showarticletitle{Approximate nearest neighbor search on high
  dimensional data -- experiments, analyses, and improvement}.
\newblock \bibinfo{journal}{\emph{IEEE TKDE}} \bibinfo{volume}{32},
  \bibinfo{number}{8} (\bibinfo{year}{2019}), \bibinfo{pages}{1475--1488}.
\newblock


\bibitem[\protect\citeauthoryear{Liang, Zhang, Yao, Chen, Song, and
  Cheng}{Liang et~al\mbox{.}}{2025}]%
        {UNIFY}
\bibfield{author}{\bibinfo{person}{Anqi Liang}, \bibinfo{person}{Pengcheng
  Zhang}, \bibinfo{person}{Bin Yao}, \bibinfo{person}{Zhongpu Chen},
  \bibinfo{person}{Yitong Song}, {and} \bibinfo{person}{Guangxu Cheng}.}
  \bibinfo{year}{2025}\natexlab{}.
\newblock \showarticletitle{UNIFY: Unified Index for Range Filtered Approximate
  Nearest Neighbors Search}.
\newblock \bibinfo{journal}{\emph{Proc. VLDB Endow.}} \bibinfo{volume}{18},
  \bibinfo{number}{4} (\bibinfo{date}{May} \bibinfo{year}{2025}),
  \bibinfo{pages}{1118–1130}.
\newblock
\showISSN{2150-8097}


\bibitem[\protect\citeauthoryear{Liu, Zeng, Chen, Ainihaer, Ramasami, Chen, Xu,
  Wu, and Wang}{Liu et~al\mbox{.}}{2025}]%
        {abs-2501-11216}
\bibfield{author}{\bibinfo{person}{Shige Liu}, \bibinfo{person}{Zhifang Zeng},
  \bibinfo{person}{Li Chen}, \bibinfo{person}{Adil Ainihaer},
  \bibinfo{person}{Arun Ramasami}, \bibinfo{person}{Songting Chen},
  \bibinfo{person}{Yu Xu}, \bibinfo{person}{Mingxi Wu}, {and}
  \bibinfo{person}{Jianguo Wang}.} \bibinfo{year}{2025}\natexlab{}.
\newblock \showarticletitle{TigerVector: Supporting Vector Search in Graph
  Databases for Advanced RAGs}.
\newblock \bibinfo{journal}{\emph{CoRR}}  \bibinfo{volume}{abs/2501.11216}
  (\bibinfo{year}{2025}).
\newblock


\bibitem[\protect\citeauthoryear{Malkov and Yashunin}{Malkov and
  Yashunin}{2018}]%
        {Hnsw}
\bibfield{author}{\bibinfo{person}{Yury Malkov} {and} \bibinfo{person}{Dmitry
  Yashunin}.} \bibinfo{year}{2018}\natexlab{}.
\newblock \showarticletitle{Efficient and robust approximate nearest neighbor
  search using hierarchical navigable small world graphs}.
\newblock \bibinfo{journal}{\emph{IEEE TPAMI}} \bibinfo{volume}{42},
  \bibinfo{number}{4} (\bibinfo{year}{2018}), \bibinfo{pages}{824--836}.
\newblock


\bibitem[\protect\citeauthoryear{Mikolov, Sutskever, Chen, Corrado, and
  Dean}{Mikolov et~al\mbox{.}}{2013}]%
        {word2vec}
\bibfield{author}{\bibinfo{person}{Tomas Mikolov}, \bibinfo{person}{Ilya
  Sutskever}, \bibinfo{person}{Kai Chen}, \bibinfo{person}{Greg~S Corrado},
  {and} \bibinfo{person}{Jeff Dean}.} \bibinfo{year}{2013}\natexlab{}.
\newblock \showarticletitle{Distributed representations of words and phrases
  and their compositionality}.
\newblock \bibinfo{journal}{\emph{NeurIPS}}  \bibinfo{volume}{26}
  (\bibinfo{year}{2013}).
\newblock


\bibitem[\protect\citeauthoryear{Nasrabadi and King}{Nasrabadi and
  King}{1988}]%
        {nasrabadi1988image}
\bibfield{author}{\bibinfo{person}{Nasser~M Nasrabadi} {and}
  \bibinfo{person}{Robert~A King}.} \bibinfo{year}{1988}\natexlab{}.
\newblock \showarticletitle{Image coding using vector quantization: A review}.
\newblock \bibinfo{journal}{\emph{IEEE Transactions on communications}}
  \bibinfo{volume}{36}, \bibinfo{number}{8} (\bibinfo{year}{1988}),
  \bibinfo{pages}{957--971}.
\newblock


\bibitem[\protect\citeauthoryear{Pan, Wang, and Li}{Pan et~al\mbox{.}}{2024}]%
        {PanWL24}
\bibfield{author}{\bibinfo{person}{James~Jie Pan}, \bibinfo{person}{Jianguo
  Wang}, {and} \bibinfo{person}{Guoliang Li}.} \bibinfo{year}{2024}\natexlab{}.
\newblock \showarticletitle{Survey of vector database management systems}.
\newblock \bibinfo{journal}{\emph{{VLDB} J.}} \bibinfo{volume}{33},
  \bibinfo{number}{5} (\bibinfo{year}{2024}), \bibinfo{pages}{1591--1615}.
\newblock


\bibitem[\protect\citeauthoryear{Patel, Kraft, Guestrin, and Zaharia}{Patel
  et~al\mbox{.}}{2024}]%
        {ACORN}
\bibfield{author}{\bibinfo{person}{Liana Patel}, \bibinfo{person}{Peter Kraft},
  \bibinfo{person}{Carlos Guestrin}, {and} \bibinfo{person}{Matei Zaharia}.}
  \bibinfo{year}{2024}\natexlab{}.
\newblock \showarticletitle{ACORN: Performant and Predicate-Agnostic Search
  Over Vector Embeddings and Structured Data}.
\newblock \bibinfo{journal}{\emph{Proceedings of the ACM on Management of
  Data}} \bibinfo{volume}{2}, \bibinfo{number}{3} (\bibinfo{year}{2024}),
  \bibinfo{pages}{1 -- 27}.
\newblock


\bibitem[\protect\citeauthoryear{Peng, Choi, Chan, Yang, and Xu}{Peng
  et~al\mbox{.}}{2023}]%
        {taumg}
\bibfield{author}{\bibinfo{person}{Yun Peng}, \bibinfo{person}{Byron Choi},
  \bibinfo{person}{Tsz~Nam Chan}, \bibinfo{person}{Jianye Yang}, {and}
  \bibinfo{person}{Jianliang Xu}.} \bibinfo{year}{2023}\natexlab{}.
\newblock \showarticletitle{Efficient Approximate Nearest Neighbor Search in
  Multi-dimensional Databases}.
\newblock \bibinfo{journal}{\emph{Proc. {ACM} Manag. Data}}
  \bibinfo{volume}{1}, \bibinfo{number}{1} (\bibinfo{year}{2023}),
  \bibinfo{pages}{54:1--54:27}.
\newblock


\bibitem[\protect\citeauthoryear{Samet}{Samet}{1984}]%
        {samet1984quadtree}
\bibfield{author}{\bibinfo{person}{Hanan Samet}.}
  \bibinfo{year}{1984}\natexlab{}.
\newblock \showarticletitle{The quadtree and related hierarchical data
  structures}.
\newblock \bibinfo{journal}{\emph{Comput. Surveys}} \bibinfo{volume}{16},
  \bibinfo{number}{2} (\bibinfo{year}{1984}), \bibinfo{pages}{187--260}.
\newblock


\bibitem[\protect\citeauthoryear{Singh, Subramanya, Krishnaswamy, and
  Simhadri}{Singh et~al\mbox{.}}{2021}]%
        {abs-2105-09613}
\bibfield{author}{\bibinfo{person}{Aditi Singh}, \bibinfo{person}{Suhas~Jayaram
  Subramanya}, \bibinfo{person}{Ravishankar Krishnaswamy}, {and}
  \bibinfo{person}{Harsha~Vardhan Simhadri}.} \bibinfo{year}{2021}\natexlab{}.
\newblock \showarticletitle{FreshDiskANN: {A} Fast and Accurate Graph-Based
  {ANN} Index for Streaming Similarity Search}.
\newblock \bibinfo{journal}{\emph{CoRR}}  \bibinfo{volume}{abs/2105.09613}
  (\bibinfo{year}{2021}).
\newblock


\bibitem[\protect\citeauthoryear{Srinivasan, Raman, Chen, Bendersky, and
  Najork}{Srinivasan et~al\mbox{.}}{2021}]%
        {WIT}
\bibfield{author}{\bibinfo{person}{Krishna Srinivasan},
  \bibinfo{person}{Karthik Raman}, \bibinfo{person}{Jiecao Chen},
  \bibinfo{person}{Michael Bendersky}, {and} \bibinfo{person}{Marc Najork}.}
  \bibinfo{year}{2021}\natexlab{}.
\newblock \showarticletitle{Wit: Wikipedia-based image text dataset for
  multimodal multilingual machine learning}. In
  \bibinfo{booktitle}{\emph{Proceedings of the 44th international ACM SIGIR
  conference on research and development in information retrieval}}.
  \bibinfo{pages}{2443--2449}.
\newblock


\bibitem[\protect\citeauthoryear{Wang, Yi, Guo, Jin, Xu, Li, Wang, Guo, Li, Xu,
  Yu, Yuan, Zou, Long, Cai, Li, Zhang, Mo, Gu, Jiang, Wei, and Xie}{Wang
  et~al\mbox{.}}{2021b}]%
        {milvus}
\bibfield{author}{\bibinfo{person}{Jianguo Wang}, \bibinfo{person}{Xiaomeng
  Yi}, \bibinfo{person}{Rentong Guo}, \bibinfo{person}{Hai Jin},
  \bibinfo{person}{Peng Xu}, \bibinfo{person}{Shengjun Li},
  \bibinfo{person}{Xiangyu Wang}, \bibinfo{person}{Xiangzhou Guo},
  \bibinfo{person}{Chengming Li}, \bibinfo{person}{Xiaohai Xu},
  \bibinfo{person}{Kun Yu}, \bibinfo{person}{Yuxing Yuan},
  \bibinfo{person}{Yinghao Zou}, \bibinfo{person}{Jiquan Long},
  \bibinfo{person}{Yudong Cai}, \bibinfo{person}{Zhenxiang Li},
  \bibinfo{person}{Zhifeng Zhang}, \bibinfo{person}{Yihua Mo},
  \bibinfo{person}{Jun Gu}, \bibinfo{person}{Ruiyi Jiang}, \bibinfo{person}{Yi
  Wei}, {and} \bibinfo{person}{Charles Xie}.} \bibinfo{year}{2021}\natexlab{b}.
\newblock \showarticletitle{Milvus: {A} Purpose-Built Vector Data Management
  System}. In \bibinfo{booktitle}{\emph{{SIGMOD} '21: International Conference
  on Management of Data}}. \bibinfo{publisher}{{ACM}},
  \bibinfo{pages}{2614--2627}.
\newblock


\bibitem[\protect\citeauthoryear{Wang, Lv, Xu, Wang, Yue, and Ni}{Wang
  et~al\mbox{.}}{2023}]%
        {NHQ}
\bibfield{author}{\bibinfo{person}{Mengzhao Wang}, \bibinfo{person}{Lingwei
  Lv}, \bibinfo{person}{Xiaoliang Xu}, \bibinfo{person}{Yuxiang Wang},
  \bibinfo{person}{Qiang Yue}, {and} \bibinfo{person}{Jiongkang Ni}.}
  \bibinfo{year}{2023}\natexlab{}.
\newblock \showarticletitle{An efficient and robust framework for approximate
  nearest neighbor search with attribute constraint}.
\newblock \bibinfo{journal}{\emph{Advances in Neural Information Processing
  Systems}}  \bibinfo{volume}{36} (\bibinfo{year}{2023}),
  \bibinfo{pages}{15738--15751}.
\newblock


\bibitem[\protect\citeauthoryear{Wang, Xu, Yue, and Wang}{Wang
  et~al\mbox{.}}{2021a}]%
        {survey2021}
\bibfield{author}{\bibinfo{person}{Mengzhao Wang}, \bibinfo{person}{Xiaoliang
  Xu}, \bibinfo{person}{Qiang Yue}, {and} \bibinfo{person}{Yuxiang Wang}.}
  \bibinfo{year}{2021}\natexlab{a}.
\newblock \showarticletitle{A Comprehensive Survey and Experimental Comparison
  of Graph-Based Approximate Nearest Neighbor Search}.
\newblock \bibinfo{journal}{\emph{PVLDB}} \bibinfo{volume}{14},
  \bibinfo{number}{11} (\bibinfo{year}{2021}), \bibinfo{pages}{1964–1978}.
\newblock


\bibitem[\protect\citeauthoryear{Wang, He, Tong, Zhou, and Zhong}{Wang
  et~al\mbox{.}}{2025}]%
        {TSANN}
\bibfield{author}{\bibinfo{person}{Yuxiang Wang}, \bibinfo{person}{Ziyuan He},
  \bibinfo{person}{Yongxin Tong}, \bibinfo{person}{Zimu Zhou}, {and}
  \bibinfo{person}{Yiman Zhong}.} \bibinfo{year}{2025}\natexlab{}.
\newblock \showarticletitle{Timestamp Approximate Nearest Neighbor Search over
  High-Dimensional Vector Data}. In \bibinfo{booktitle}{\emph{ICDE}}. IEEE
  Computer Society, \bibinfo{pages}{3043--3055}.
\newblock


\bibitem[\protect\citeauthoryear{Wei, Wu, Wang, Lou, Zhan, Li, and Cai}{Wei
  et~al\mbox{.}}{2020}]%
        {ADBV}
\bibfield{author}{\bibinfo{person}{Chuangxian Wei}, \bibinfo{person}{Bin Wu},
  \bibinfo{person}{Sheng Wang}, \bibinfo{person}{Renjie Lou},
  \bibinfo{person}{Chaoqun Zhan}, \bibinfo{person}{Feifei Li}, {and}
  \bibinfo{person}{Yuanzhe Cai}.} \bibinfo{year}{2020}\natexlab{}.
\newblock \showarticletitle{AnalyticDB-V: {A} Hybrid Analytical Engine Towards
  Query Fusion for Structured and Unstructured Data}.
\newblock \bibinfo{journal}{\emph{PVLDB}} \bibinfo{volume}{13},
  \bibinfo{number}{12} (\bibinfo{year}{2020}), \bibinfo{pages}{3152--3165}.
\newblock


\bibitem[\protect\citeauthoryear{Xie, Yu, and Liu}{Xie et~al\mbox{.}}{2025a}]%
        {XieYL25}
\bibfield{author}{\bibinfo{person}{Jiadong Xie}, \bibinfo{person}{Jeffrey~Xu
  Yu}, {and} \bibinfo{person}{Yingfan Liu}.} \bibinfo{year}{2025}\natexlab{a}.
\newblock \showarticletitle{Fast Approximate Similarity Join in Vector
  Databases}.
\newblock \bibinfo{journal}{\emph{Proc. {ACM} Manag. Data}}
  \bibinfo{volume}{3}, \bibinfo{number}{3} (\bibinfo{year}{2025}),
  \bibinfo{pages}{158:1--158:26}.
\newblock


\bibitem[\protect\citeauthoryear{Xie, Yu, and Liu}{Xie et~al\mbox{.}}{2025b}]%
        {ALMG}
\bibfield{author}{\bibinfo{person}{Jiadong Xie}, \bibinfo{person}{Jeffrey~Xu
  Yu}, {and} \bibinfo{person}{Yingfan Liu}.} \bibinfo{year}{2025}\natexlab{b}.
\newblock \showarticletitle{Graph Based K-Nearest Neighbor Search Revisited}.
\newblock \bibinfo{journal}{\emph{ACM Trans. Database Syst.}}
  (\bibinfo{date}{May} \bibinfo{year}{2025}).
\newblock


\bibitem[\protect\citeauthoryear{Xie, Yu, Teng, and Liu}{Xie
  et~al\mbox{.}}{2025c}]%
        {XieYTL25}
\bibfield{author}{\bibinfo{person}{Jiadong Xie}, \bibinfo{person}{Jeffrey~Xu
  Yu}, \bibinfo{person}{Siyi Teng}, {and} \bibinfo{person}{Yingfan Liu}.}
  \bibinfo{year}{2025}\natexlab{c}.
\newblock \showarticletitle{Beyond Vector Search: Querying With and Without
  Predicates}.
\newblock \bibinfo{journal}{\emph{Proc. {ACM} Manag. Data}}
  \bibinfo{volume}{3}, \bibinfo{number}{6} (\bibinfo{year}{2025}),
  \bibinfo{pages}{1--26}.
\newblock


\bibitem[\protect\citeauthoryear{Xu, Gao, Gou, Long, and Jensen}{Xu
  et~al\mbox{.}}{2025}]%
        {iRange}
\bibfield{author}{\bibinfo{person}{Yuexuan Xu}, \bibinfo{person}{Jianyang Gao},
  \bibinfo{person}{Yutong Gou}, \bibinfo{person}{Cheng Long}, {and}
  \bibinfo{person}{Christian~S Jensen}.} \bibinfo{year}{2025}\natexlab{}.
\newblock \showarticletitle{iRangeGraph: Improvising Range-dedicated Graphs for
  Range-filtering Nearest Neighbor Search}.
\newblock \bibinfo{journal}{\emph{Proceedings of the ACM on Management of
  Data}} \bibinfo{volume}{2}, \bibinfo{number}{6} (\bibinfo{year}{2025}),
  \bibinfo{pages}{1--26}.
\newblock


\bibitem[\protect\citeauthoryear{Yang, Cai, and Zheng}{Yang
  et~al\mbox{.}}{2024}]%
        {YangCZ24}
\bibfield{author}{\bibinfo{person}{Ming Yang}, \bibinfo{person}{Yuzheng Cai},
  {and} \bibinfo{person}{Weiguo Zheng}.} \bibinfo{year}{2024}\natexlab{}.
\newblock \showarticletitle{{CSPG:} Crossing Sparse Proximity Graphs for
  Approximate Nearest Neighbor Search}. In \bibinfo{booktitle}{\emph{NeurIPS
  2024}}.
\newblock


\bibitem[\protect\citeauthoryear{Yang, Cai, and Zheng}{Yang
  et~al\mbox{.}}{2025a}]%
        {Hi-PNG}
\bibfield{author}{\bibinfo{person}{Ming Yang}, \bibinfo{person}{Yuzheng Cai},
  {and} \bibinfo{person}{Weiguo Zheng}.} \bibinfo{year}{2025}\natexlab{a}.
\newblock \showarticletitle{Hi-PNG: Efficient Interval-Filtering ANNS via
  Hierarchical Interval Partition Navigating Graph}. In
  \bibinfo{booktitle}{\emph{SIGKDD}}. \bibinfo{pages}{3518--3529}.
\newblock


\bibitem[\protect\citeauthoryear{Yang, Xie, Liu, Yu, Gao, Wang, Peng, and
  Cui}{Yang et~al\mbox{.}}{2025b}]%
        {FastPG}
\bibfield{author}{\bibinfo{person}{Shuo Yang}, \bibinfo{person}{Jiadong Xie},
  \bibinfo{person}{Yingfan Liu}, \bibinfo{person}{Jeffrey~Xu Yu},
  \bibinfo{person}{Xiyue Gao}, \bibinfo{person}{Qianru Wang},
  \bibinfo{person}{Yanguo Peng}, {and} \bibinfo{person}{Jiangtao Cui}.}
  \bibinfo{year}{2025}\natexlab{b}.
\newblock \showarticletitle{Revisiting the Index Construction of Proximity
  Graph-Based Approximate Nearest Neighbor Search}.
\newblock \bibinfo{journal}{\emph{PVLDB}} \bibinfo{volume}{18},
  \bibinfo{number}{6} (\bibinfo{year}{2025}), \bibinfo{pages}{1825--1838}.
\newblock


\bibitem[\protect\citeauthoryear{Yu, Lin, Gong, Xie, Liu, Zhou, Sun, Zhang, Li,
  and Yu}{Yu et~al\mbox{.}}{2025}]%
        {abs-2503-00402}
\bibfield{author}{\bibinfo{person}{Song Yu}, \bibinfo{person}{Shengyuan Lin},
  \bibinfo{person}{Shufeng Gong}, \bibinfo{person}{Yongqing Xie},
  \bibinfo{person}{Ruicheng Liu}, \bibinfo{person}{Yijie Zhou},
  \bibinfo{person}{Ji Sun}, \bibinfo{person}{Yanfeng Zhang},
  \bibinfo{person}{Guoliang Li}, {and} \bibinfo{person}{Ge Yu}.}
  \bibinfo{year}{2025}\natexlab{}.
\newblock \showarticletitle{A Topology-Aware Localized Update Strategy for
  Graph-Based {ANN} Index}.
\newblock \bibinfo{journal}{\emph{CoRR}}  \bibinfo{volume}{abs/2503.00402}
  (\bibinfo{year}{2025}).
\newblock


\bibitem[\protect\citeauthoryear{Zhang, Xu, Chen, Sui, Xie, Cai, Chen, He,
  Yang, Yang, Yang, and Zhou}{Zhang et~al\mbox{.}}{2023}]%
        {vbase}
\bibfield{author}{\bibinfo{person}{Qianxi Zhang}, \bibinfo{person}{Shuotao Xu},
  \bibinfo{person}{Qi Chen}, \bibinfo{person}{Guoxin Sui},
  \bibinfo{person}{Jiadong Xie}, \bibinfo{person}{Zhizhen Cai},
  \bibinfo{person}{Yaoqi Chen}, \bibinfo{person}{Yinxuan He},
  \bibinfo{person}{Yuqing Yang}, \bibinfo{person}{Fan Yang},
  \bibinfo{person}{Mao Yang}, {and} \bibinfo{person}{Lidong Zhou}.}
  \bibinfo{year}{2023}\natexlab{}.
\newblock \showarticletitle{{VBASE:} Unifying Online Vector Similarity Search
  and Relational Queries via Relaxed Monotonicity}. In
  \bibinfo{booktitle}{\emph{17th {USENIX} Symposium on Operating Systems Design
  and Implementation, {OSDI} 2023}}. \bibinfo{publisher}{{USENIX} Association},
  \bibinfo{pages}{377--395}.
\newblock


\bibitem[\protect\citeauthoryear{Zhu, Chen, Gao, Ma, Zheng, and Zhao}{Zhu
  et~al\mbox{.}}{2024}]%
        {ZhuCGMZZ24}
\bibfield{author}{\bibinfo{person}{Yifan Zhu}, \bibinfo{person}{Lu Chen},
  \bibinfo{person}{Yunjun Gao}, \bibinfo{person}{Ruiyao Ma},
  \bibinfo{person}{Baihua Zheng}, {and} \bibinfo{person}{Jingwen Zhao}.}
  \bibinfo{year}{2024}\natexlab{}.
\newblock \showarticletitle{{HJG:} An Effective Hierarchical Joint Graph for
  {ANNS} in Multi-Metric Spaces}. In \bibinfo{booktitle}{\emph{{ICDE}}}.
  \bibinfo{publisher}{{IEEE}}, \bibinfo{pages}{4275--4287}.
\newblock


\bibitem[\protect\citeauthoryear{Zuo, Qiao, Zhou, Li, and Deng}{Zuo
  et~al\mbox{.}}{2024}]%
        {SeRF}
\bibfield{author}{\bibinfo{person}{Chaoji Zuo}, \bibinfo{person}{Miao Qiao},
  \bibinfo{person}{Wenchao Zhou}, \bibinfo{person}{Feifei Li}, {and}
  \bibinfo{person}{Dong Deng}.} \bibinfo{year}{2024}\natexlab{}.
\newblock \showarticletitle{SeRF: Segment Graph for Range-Filtering Approximate
  Nearest Neighbor Search}.
\newblock \bibinfo{journal}{\emph{Proceedings of the ACM on Management of
  Data}} \bibinfo{volume}{2}, \bibinfo{number}{1} (\bibinfo{year}{2024}),
  \bibinfo{pages}{1 -- 26}.
\newblock


\end{thebibliography}


\numberwithin{theorem}{section}

\appendix

\setcounter{section}{0}
\renewcommand{\thesection}{\Alph{section}}
\renewcommand{\thesubsection}{\thesection.\arabic{subsection}}

\numberwithin{theorem}{section}
\numberwithin{lemma}{section}

\section{Atomic RR Predicates vs Base Relations of Allen’s Interval Algebra}
\label{sec:correspondence}

Allen’s Interval Algebra defines a total of 13 base relations between two intervals/ranges. Let $X = [l_q, r_q]$ and $Y = [l_i, r_i]$. We show that 11 of those 13 relations could be represented by our four atomic RR predicates as shown in Fig.~\ref{fig:illus_rf}. 

\begin{itemize}
    \item $X \ \mathbf{m} \  Y \iff l_q\le r_q=l_i\le r_i$ is a special case of \textcircled{3}
    \item $X \ \mathbf{mi} \  Y \iff l_i\le r_i=l_q\le r_q$ is a special case of \textcircled{1}
    \item $X \ \mathbf{o} \  Y \iff l_q < l_i < r_q < r_i$ is a special case of \textcircled{3}
    \item $X \ \mathbf{oi} \  Y \iff l_i < l_q < r_i < r_q$ is a special case of \textcircled{1}
    \item $X \ \mathbf{s} \  Y \iff l_q = l_i < r_q < r_i$ is a special case of \textcircled{2}
    \item $X \ \mathbf{si} \  Y \iff l_i = l_q < r_i < r_q$ is a special case of \textcircled{4}
    \item $X \ \mathbf{d} \  Y \iff l_i < l_q < r_q < r_i$ is a special case of \textcircled{2}
    \item $X \ \mathbf{di} \  Y \iff l_q < l_i < r_i < r_q$ is a special case of \textcircled{4}
    \item $X \ \mathbf{f} \  Y \iff l_i < l_q < r_q = r_i$ is a special case of \textcircled{2}
    \item $X \ \mathbf{fi} \  Y \iff l_q < l_i < r_i = r_q$ is a special case of \textcircled{4}
    \item $X \ \mathbf{=} \  Y \iff l_i = l_q < r_i = r_q$ is a special case of \textcircled{2}
\end{itemize}

The remaining two relations $<,>$ mean no intersections between $X$ and $Y$, which could not be represented by our four atomic RR predicates. But, they can still be supported by MSTG. To be specific, $X~<~Y$ requires $l_q \le r_q < l_i \le r_i$, which could be reduced to the RFANN filter $r_q < l_i$ since $l_q \le r_q$ and $l_i \le r_i$ holds according to the definitions of the object range and query range. Similarly, $Y~<~X$ is reduced to  $r_i < l_q$. Since our method is able to solve RFANN queries, MSTG could address the filters of both $X~<~Y$ and $Y~<~X$.  

\section{Details of Algorithms}
\label{sec:details_alg}

\stitle{Search Algorithm over A PG:} As shown in Algorithm~\ref{alg:knn_search}, the search process starts from an entering point $ep$ and puts it in a sorted array $pool$ of nodes, which is maintained to store the currently found $L$-closest neighbors (lines 1-2). Then, it iteratively extracts the closest but unexpanded neighbor $u$ from $pool$ (line 4) and expands $u$ to refine $pool$, until the termination condition is satisfied (line 3).
In each iteration, expanding $u$ for $q$ is shown in Lines 5-7, where each neighbor $v \in N_G(u)$ is treated as a $k$-ANN candidate of $q$ (line 5) and further verified by an expensive distance computation  (line 6) to refine $pool$ (line 7). At the end of each iteration (line 8), the algorithm finds the closest but unexpanded vertex in $pool$ as the next one to be expanded. It terminates when the first $L$ vertices in $pool$ have been expanded (line 3).

\begin{algorithm}[t]
\small  
    \SetVline 
    \SetFuncSty{textsf}
    \SetArgSty{textsf}
 \caption{\texttt{KANNSearch}($G, q, k, L, ep$)}
 \label{alg:knn_search}
\Input{PG $G$, query $q$, $k$, pool width $L$ and entering point $ep$}
 \Output{$k$-ANN of query point $q$}
\State{$i \leftarrow 0$}
\State{$pool[0] \leftarrow (ep,dist(q, ep))$}
\While{$i < L$} 
{
	\State{$u\leftarrow pool[i]$}
	\For{each {$v \in N_G(u)$}}
        {
		\State{insert $(v, dist(q, v))$ into $pool$}
	}
	\State{sort $pool$ and keep the $L$ closest neighbors}
	\State{$i \leftarrow $ index of the first unexpanded vertex in $pool$}
}
\Return{$pool[0, \ldots, k-1]$}
\end{algorithm}

\section{Details of General-Purpose Approaches}
\label{sec:details_gpa}

General-purpose methods, which support $k$-ANNS with arbitrary filters, including pre-filtering~\cite{ADBV,milvus,vbase}, post-filtering~\cite{ADBV,milvus}, Milvus~\cite{milvus}, VBASE~\cite{vbase}, and ACORN~\cite{ACORN}.

Pre-filtering~\cite{ADBV,milvus,vbase} involves initially retrieving a subset of objects that satisfy the query predicate and then generating the results on this subset by the brute-force scan. 
It is easy to implement, but it is only efficient for low-selectivity filters.
Conversely, post-filtering~\cite{ADBV,milvus} first performs $k$-ANNS on the entire set of objects and returns $k^{\prime}$ ($k^{\prime} \geq k$) objects that are subsequently verified by the query filter. However, it is challenging to determine the appropriate $k^{\prime}$. A small $k^{\prime}$ value may result in an insufficient number of qualified results returned, while a large $k^{\prime}$ value decreases the search efficiency.
Therefore, Milvus~\cite{milvus} determines $k^{\prime}$ in a progressive manner, where $k^{\prime}$ starts from $k$ and then is doubled until $\geq k$ qualified results can be returned after the filtering on the $k^{\prime}$-ANN obtained.
However, Milvus is still inefficient for multiple $k^{\prime}$-ANNS.
VBASE~\cite{vbase} introduces a new search algorithm with two phases. First, it ignores the query predicate and directs the search towards the $1$-ANN of the query vector. Second, it considers the query predicate, i.e., greedily traverses the nodes that satisfy the query predicate on PG as in Algorithm~\ref{alg:knn_search}.
Moreover, ACORN~\cite{ACORN} employs a graph index with enlarged node out-degree, i.e., considering 2-hop neighbors as neighbors in the new graph for each node.
During the search process, ACORN follows the procedure of Algorithm~\ref{alg:knn_search}, but only traverses the neighbors in $N_G(u)$ that satisfy the query predicate in line 5 of Algorithm~\ref{alg:knn_search}.

\section{Lemmas, Theorems and Proofs}
\label{sec:proofs}

\begin{lemma}
Consider a segment tree $T$ constructed based on elements from a numeric attribute $A$. For any arbitrary range $[l_q,r_q]$, let $p$ be the smallest value such that there exist $p$ nodes in $T$ where the union range indicated by these $p$ nodes is $\{a_i\in A\mid l_q\le a_i\le r_q\}$. The value of $p$ is bounded by $O(\log |A|)$.
\label{lemma:log-segment-tree}
\end{lemma}

\proofsketch 
For each arbitrary range $[l_q,r_q]$ as a query, we can initiate the search for these nodes from the root node and apply recursion to its two child nodes whenever the node only partially overlaps (i.e., is not entirely contained) within the query range. 
At each level of the segment tree, a maximum of two nodes are partially overlapping with the query range, necessitating additional recursion. Given that the depth of the tree is $O(\log |A|)$, the number of such nodes, denoted as $p$, is constrained by $O(\log |A|)$.
\eop

\begin{theorem}
For a given $x\in [1,|A|]$, consider a tree node $\mathcal{T}^x_{l,r}$ in MSTG. Let $G_x$ be the induced subgraph, where the label $(b,e)$ of each edge in $G_x$ satisfying $x \in [b,e]$, from the tree node $\mathcal{T}^{|A|}_{l,r}$ in the labeled MSTG. 
Under the same parameters $ef_{con}$ and $m$, the PG $\mathcal{T}^x_{l,r}.G$ in MSTG is identical to the induced $G_x$ in the labeled MSTG.
\label{theo:mstg-same}
\end{theorem}

\proofsketch 
Firstly, when $x=1$, the edges in $G_1$ are the edges of the HNSW containing objects $\{o_i=(v_i,l_i,r_i)\mid l_i=a_1 \wedge r_i\in [l,r]\}$, since no other edges' range include $1$. Hence, $G_1$ is identical to $\mathcal{T}^1_{l,r}.G$.
Assume that for $x=y\in [1,n)$, we have $G_y$ is identical to $\mathcal{T}^y_{l,r}.G$, we prove below that when $x=y+1$, we have $G_x$ is identical to PG in $\mathcal{T}^x_{l,r}.G$.
Let $G'=G_{x}$ be the HNSW graph being constructed before inserting objects $O'_{x+1}=\{o_i=(v_i,l_i,r_i)\in O\mid l_i=a_{x+1} \wedge r_i\in [l,r]\}$. Compare the graph $G'$ and the PG $\mathcal{T}_{l,r}^{x+1}.G$, the differences lie in the additional edges connecting nodes of $O'_{x+1}$, and the removal due to the RNG pruning strategy. Considering the insertion of objects in $O'_{x+1}$ one by one into $G_x$ in Algorithm~\ref{alg:build_pg} to obtain $G_{x+1}$, the edges from each inserted objects in $O_{x+1}$ is labeled with $[x+1,+\infty]$ (line 5) and the edges pruned with labeled with $[b,x]$ (line 10). Hence, the labeled range of inserted edges included $x+1$ and the removed edges excluded $x+1$, which leads to $G_{x+1}$ being identical to the PG $\mathcal{T}^{x+1}_{l,r}.G$. This completes the induction.
\eop

\stitle{Proof Sketch of Theorem~\ref{theo:combine-predicate}:}
We first discuss the cases of disjunction between two conditions.

\stitle{$R_q$ is \textcircled{1}$\vee$\textcircled{2}}: Since $l_q\le r_q$ and $l_i\le r_i$ always hold, the conditions $(l_i\leq l_q\leq r_i\leq r_q) \vee (l_i\leq l_q\leq r_q\leq r_i) \Leftrightarrow (l_i\leq l_q)\wedge(( l_q\leq r_i\leq r_q )\vee(r_q\leq r_i))\Leftrightarrow (l_i\leq l_q)\wedge(r_i\ge l_q)$.
Therefore, we can utilize a single MSTG $\mathcal{T}$ to answer to a query $q=(v_q,l_q,r_q)$ by identifying a value of $x$ where $a_x\leq l_q< a_{x+1}$ (assuming $a_{|A|+1}=+\infty$), and querying the range $[l_q,+\infty]$ in the segment tree of $\mathcal{T}^x$.

\stitle{$R_q$ is \textcircled{2}$\vee$\textcircled{3}}: 
Since $l_q\le r_q$ and $l_i\le r_i$ always hold, the conditions $(l_i\leq l_q\leq r_q\leq r_i)\vee (l_q\leq l_i\leq r_q\leq r_i)\Leftrightarrow ((l_i\le l_q)\vee (l_q\le l_i\le r_q)) \wedge (r_q\le r_i)\Leftrightarrow (l_i\le r_q)\wedge  (r_q\le r_i)$.
Therefore, we can utilize a single MSTG $\mathcal{T}$ to answer to a query $q=(v_q,l_q,r_q)$ by identifying a value of $x$ where $a_{x} \leq r_q < a_{x+1}$ (assuming $a_{|A|+1}=+\infty$), and querying the range $[r_q, +\infty]$ in the segment tree of $\mathcal{T}^x$.

\stitle{$R_q$ is \textcircled{3}$\vee$\textcircled{4}}:
Since $l_q\le r_q$ and $l_i\le r_i$ always hold, the conditions $(l_q\leq l_i\leq r_q\leq r_i)\vee(l_q\leq l_i\leq r_i\leq r_q)\Leftrightarrow (l_q\leq l_i\leq r_q)\wedge ((r_q\leq r_i)\vee (l_q\leq r_i\leq r_q))\Leftrightarrow (l_q\leq l_i\leq r_q)\wedge (r_i\geq l_q)$.
Therefore, we can utilize a single MSTG $\mathcal{T}'$ to answer to a query $q=(v_q,l_q,r_q)$ by identifying a value of $x$ where $a_{x-1}< l_q\leq a_{x}$ (assuming $a_{0}=-\infty$), and querying the range $[l_q,r_q]$ in the segment tree of $\mathcal{T}'^x$.

Therefore, the above three RR predicates, the disjunctions of two cases, only require one MSTG. This means we need at most two MSTG indexes to address the RR predicates of (1) the disjunction of any three cases or (2) the disjunction of four cases. For example, when $R_q$ is \textcircled{1}$\vee$\textcircled{2}$\vee$\textcircled{3}$\vee$\textcircled{4}, we can merge the results of RRANN queries with $R_q$ as \textcircled{1}$\vee$\textcircled{2} and $R_q$ as \textcircled{3}$\vee$\textcircled{4} respectively to derive the final results.

In conclusion, a RRANN query with any combined RR predicates requires at most two MSTG indexes and at most two separate searches to merge their results for the final result.\eop

\section{Experimental Settings}
\label{sec:exp_setting}

\begin{table}[H]
\centering
\caption{Statistics of datasets}
\vspace{-3mm}
\label{tb:data}
\resizebox{0.85\linewidth}{!}{
\begin{tabular}{|c|r|r|r|c|}
\hline
\textbf{Dataset} &  \textbf{\#vectors} &  \textbf{\#queries} & \textbf{dim.}& \textbf{type} \\
\hline \hline
Sift   & 1,000,000  & 10,000 & 128& Image \\
 \hline
Gist   & 1,000,000& 1000  & 960 & Image \\
 \hline
WIT-Image      & 1,000,000 & 1,000 & 2,048 & Image \\
 \hline
Paper    & 1,000,000& 10,000 & 200   & Text \\
 \hline
Redcaps  & 1,000,000& 1,000  & 512  & Image \& Text \\
\hline
Sift50M   & 50,000,000  & 10,000 & 128& Image \\
 \hline
\end{tabular}
}
\end{table} \vspace{-3mm}

\noindent \textbf{Details of Datasets:}
The data statistics are summarized in Table \ref{tb:data}, where \emph{\#vectors} represents the dataset size, \emph{\#queries} the number of queries, and \emph{dim.} the vector dimensionality. 
Sift50M, sampled from Sift1B~\cite{sift}, is utilized to test the scalability of MSTG.
For Sift, Gist and Paper, the original query vectors are provided. For Redcaps, 1,000 query vectors are generated by prompting ChatGPT-4 to create queries for an image search system and embedding them using CLIP. 
For WIT-Image, as in ~\cite{iRange}, 1,000 query vectors are randomly sampled from the dataset. 
As to the attribute ranges, we assign the vectors the ranges in $[0, 10^4)$ with various distributions, including uniform, long-tail, normal, Poisson, and Zipf, where uniform is the default setting unless specified. 
As to the single attribute value in RFANN and TSANN, we assign numerical encodings of categorical attributes, image sizes and timestamps to objects in Paper, WIT-Image and Redcaps respectively, while randomly generated values in Sift and Gist as in~\cite{iRange,UNIFY}.
The query ranges are randomly determined according to the specified selectivity, i.e., the ratio of objects satisfying the query filter.

\noindent \textbf{Compared Algorithms and Parameters}:
We first compare our approach \textbf{MSTG}, with RRANN methods in Section~\ref{sec:limit}.
For each method, we use recommended or default parameters if provided. Otherwise, we tune them for the best performance.
\textbf{(1) ACORN~\cite{ACORN}}: ACORN has two versions: \textbf{ACORN-$\gamma$} and \textbf{ACORN-$1$}, which enlarge the neighbor list of each node in index construction and search phases, respectively.
For ACORN-$\gamma$, we set $M = 32$ and $M_\beta = 64$, where $\gamma=12$ obtained through grid search. Other parameters retain the default ones. ACORN-$1$ shares the same parameters as ACORN-$\gamma$ except $\gamma = 1$.
\textbf{(2) Post-filtering~\cite{ADBV,milvus}}: Post-filtering first retrieves $k^{'}$-ANN ($k^{'} \ge k$) with HNSW, and then applies filtering to them to derive the final $k$ results. We set $M = 16$ and $ef_{con} = 200$ to build the HNSW index via a grid search. $k^{'}$ is also selected by grid search for each dataset. 
\textbf{(3) Pre-filtering~\cite{ADBV,milvus,vbase}}: 
No parameter needs to be tuned for pre-filtering.
\textbf{(4) Milvus~\cite{milvus}}: Milvus partitions the dataset based on attribute value ranges and employs a cost model to select between pre-filtering and post-filtering for each subset. It uses HNSW as the vector index and \texttt{STL\_SORT} as the scalar index to accelerate the filtering. For fairness, its parameters are aligned with post-filtering.
VBASE~\cite{vbase} is excluded from our comparison due to its poor performance~\cite{WST,iRange}.
FilteredVamana~\cite{Filtered-diskann}, StitchedVamana~\cite{Filtered-diskann}, and UNG~\cite{UNG} are also excluded. Specifically, FilteredVamana and UNG suffer from huge construction costs caused by $10^4$ labels used in our experiments. UNG efficiently supports only a few dozen labels~\cite{UNG}. As to StitchedVamana, its construction frequently encounters out-of-memory errors.

Since RFANN is a special case of RRANN, our method naturally supports RFANN queries and thus we compare MSTG with the SOTA RFANN methods.
\textbf{(5) 2DSegmentGraph~\cite{SeRF}}: 
As recommended in~\cite{SeRF}, we set $M = 64$ and $K = 100$ for WIT-Image. Following~\cite{iRange}, we set $M = 32$ and $K = 100$ for Redcaps. For the rest, we use $M = 16$ and $K = 200$ via a grid search.
\textbf{(6) \textbf{HSIG}~\cite{UNIFY}}: HSIG samples a subset of objects and constructs PGs over partitioned data to facilitate RFANN queries. According to ~\cite{UNIFY}, we determine $S = 8$, $M = 16$, and $ef_{con} = 500$, retaining default values for other parameters.
\textbf{(7)  \textbf{SuperPostfiltering}~\cite{WST}}: SuperPostfiltering first determines multiple overlapping ranges and establishes a graph index for each range. During search, it selects the smallest covering range and applies post-filtering on the corresponding index. As per~\cite{WST}, we set $m = 64$, $EF = 500$, and $\beta = 2$ for all datasets.
\textbf{(8) \textbf{iRangeGraph}~\cite{iRange}}: 
As in~\cite{iRange}, we set $M = 64$ and $ef_{con} = 400$ for RedCaps, $M = 64$ and $ef_{con} = 100$ for WIT-Image, and $M = 16$ and $ef_{con} = 200$ for the remaining ones based on grid search.

As TSANN and IFANN are two special cases of RRANN, we compare MSTG with \textbf{(9) TS-Graph}~\cite{TSANN} for TSANN queries and \textbf{(10) Hi-PNG}~\cite{Hi-PNG} for IFANN queries, respectively. 
Following the parameter settings outlined in their paper, we configured $M=16, M'=200$, and $\mu=8$ for TS-Graph, and $M=32, ef_{con} =128$ for Hi-PNG. We opt for HNSW as the PG index for Hi-PNG due to its best performance as shown in \cite{Hi-PNG}.

Our approach, \textbf{(11) MSTG}, has two parameters in the construction of HNSW for each segment tree node, i.e., (1) $M$ that defines the maximum out-degree per node, and (2) $ef_{con}$ that specifies the size of the candidate neighbor list. Through a grid search, we set $M = 32$ and $ef_{con} = 200$ for all range filtering queries, i.e., RRANN, RFANN, IFANN, and TSANN.

\noindent \textbf{Computing Environment}:
{The experiments are conducted on a server equipped with an Intel(R) Xeon(R) CPU E5-2682 v4 CPU @2.50GHz and {64 GB DRAM}, whose OS version is Ubuntu 22.04 LTS, except the scalability test (Exp. 13) that is run on the server with two Intel Xeon Gold 6238 @ 2.10GHz and 1TB DRAM. All codes were written in C++ and compiled by {g++ 11.4.0} with {-O3} flag. SIMD instructions are enabled to accelerate distance computations. The index construction uses 16 threads, while the search performance is evaluated using a single thread.}

\section{Results of Other Experiments}
\label{sec:exp_others}

\begin{figure*}[t]
  \centering
  \includegraphics[width=0.5\linewidth]{figures/interval_exp/interval_filter_qps_vs_recall_multi_panel_legend.pdf}
    \includegraphics[width=\linewidth]{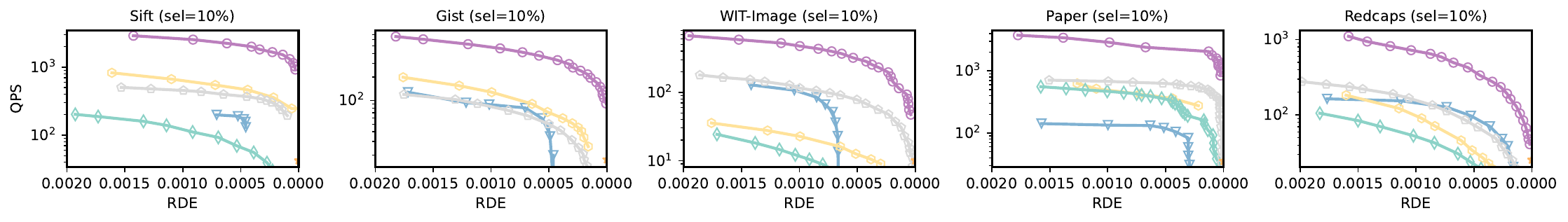}
  \vspace{-6mm}
  \caption{Overall query performance of RRANN with RDE as the accuracy measure (Exp. 1)}
  \label{fig:overall_interval_rde}
  \vspace{-2mm}
\end{figure*}

\noindent \textbf{{Exp. 1}: Overall query performance of RRANN Queries.}
As shown in Fig.~\ref{fig:overall_interval_rde}, our method MSTG significantly outperforms all its competitors for RRANN queries, with RDE as the accuracy measure. This result is consistent with that in Fig.~\ref{fig:overall_interval}, where $Recall@k$ is employed as the accuracy measure. 

\begin{figure}[H]
  \centering
    \includegraphics[width=\linewidth]{figures/interval_exp/interval_filter_qps_vs_recall_multi_panel_legend.pdf}
  
  \includegraphics[width=\linewidth]{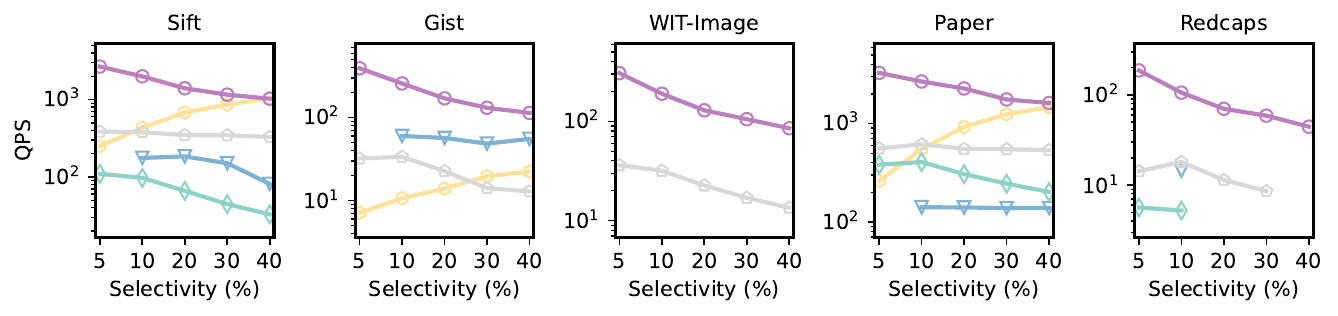}
  \vspace{-5mm}
  \caption{Impact of query selectivity of RRANN (Exp. 7)}
  \label{fig:overall_vary_sel_recall0.99}
\end{figure}

\noindent \textbf{{Exp. 7}: Impact of Query Selectivity.}
As shown in Fig.~\ref{fig:overall_vary_sel_recall0.99}, we vary the query selectivity across $\{5\%, 10\%, 20\%, 30\%, 40\%\}$ and report QPS at Recall@10 at 0.99.
We exclude the methods that fail to achieve Recall@k=0.99 and QPS<10.
We can see that MSTG outperforms its competitors in various selectivity levels. Moreover, its efficiency decreases as the selectivity increases. 
Notably, the performance of Post-filtering improves as selectivity rises, especially on Sift and Paper, because each member of $k^{'}$-ANN has a growing probability of passing the filter, which thus reduces the unnecessary exploration. 

\begin{figure*}[t]
  \centering
  \includegraphics[width=0.5\linewidth]{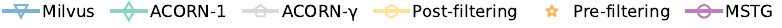}

  \includegraphics[width=\linewidth]{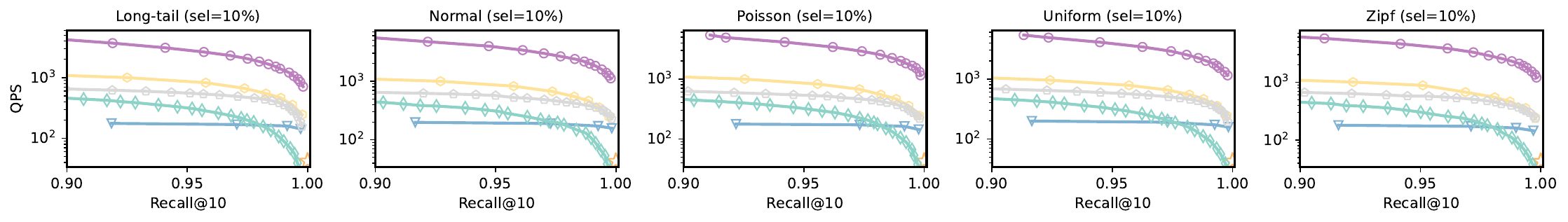}
  \includegraphics[width=\linewidth]{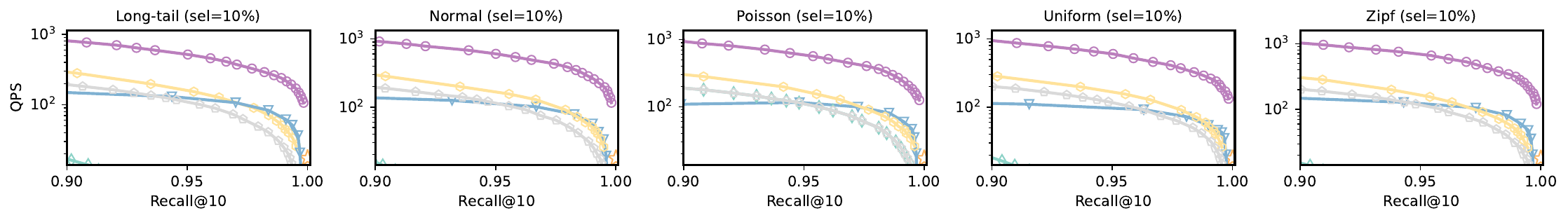}
  \vspace{-4mm}
  \caption{The impact of attribute distribution on RRANN search performance (Exp. 8). Two lines represent the results of Sift and Gist, respectively.} 
  \label{fig:distribution}
  \vspace{-3mm}
\end{figure*}

\noindent \textbf{Exp. 8: Impact of Attribute Distribution}
We vary the attribute distributions of datasets, and present the effect of them on RRANN search performance in Fig.~\ref{fig:distribution}, where we employ five different distributions, i.e., long-tail, normal, Poisson, uniform, and Zipf, to generate the attribute values. From the results, our method, MSTG, consistently outperforms its competitors in various distributions.

\begin{figure}[H]
  \centering
    \includegraphics[width=0.35\linewidth]{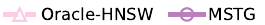}
  
  \includegraphics[width=\linewidth]{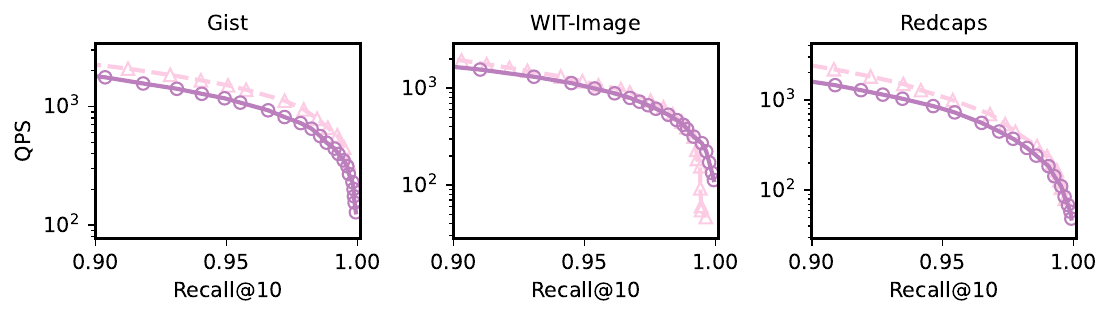}
  \vspace{-5mm}
  \caption{MSTG vs. Oracle-HNSW (Exp. 9)}
  \label{fig:oracle_interval_sel5}
  \vspace{-3mm}
\end{figure}

\noindent \textbf{Exp. 9: MSTG vs. Oracle-HNSW.}
As aforementioned, our primary goal is to achieve search performance comparable to $k$-ANN search on a PG. Here, we compare MSTG with Oracle-HNSW, where a specific HNSW is constructed for each query to manage the vectors satisfying the query filter.
Note that Oracle-HNSW is not a practical solution, because building such an HNSW for each query is unfeasible due to the impossibility of knowing the query in advance.
As shown in Fig.~\ref{fig:oracle_interval_sel5}, MSTG achieves performance comparable to Oracle-HNSW when the query selectivity is 5\% on Gist, WIT-Image, and Redcaps. 
This is because MSTG builds the PG for the vectors satisfying the query filter in an online and virtual manner, resulting in a slightly extra cost during search. 
Similar phenomena could be observed on other datasets and selectivity values, which are omitted due to space limitations. 

\begin{figure}[H]
  \centering
  
  \includegraphics[width=0.3\linewidth]{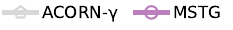}
  
  \includegraphics[width=\linewidth]{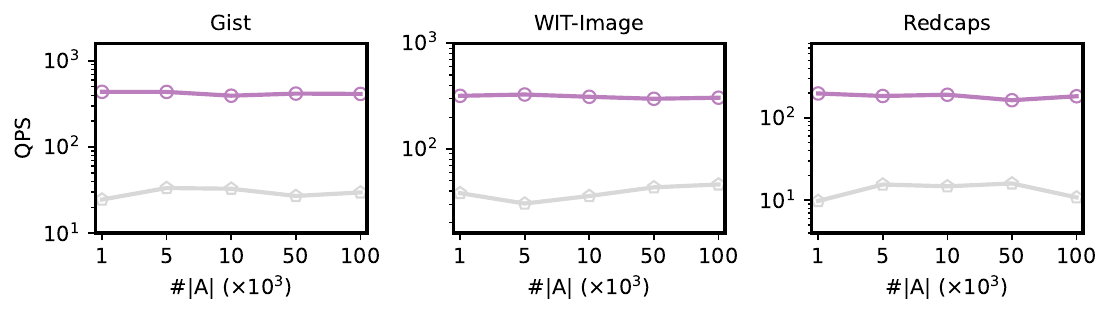}
  \vspace{-5mm}
  \caption{Impact of the Cardinality of $A$ (Exp. 10)}
  \label{fig:interval_vary_base_range_sel5}
  \vspace{-3mm}
\end{figure}

\noindent \textbf{Exp. 10: Impact of the Cardinality of $A$.}
We vary the size of the attribute $A$ from $10^3$ and $10^5$ to show its effect on the RRANN performance.
As shown in Fig.~\ref{fig:interval_vary_base_range_sel5}, where the query selectivity is 5\% and Recall@10 = 0.99, both MSTG and ACORN-$\gamma$ maintain stable QPS for various $|A|$ values. We exclude other methods, since their QPS values are below 10 in this setting.

\begin{figure}[H]
  \centering
   \includegraphics[width=0.7\linewidth]{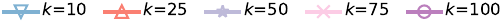}
  \includegraphics[width=\linewidth]{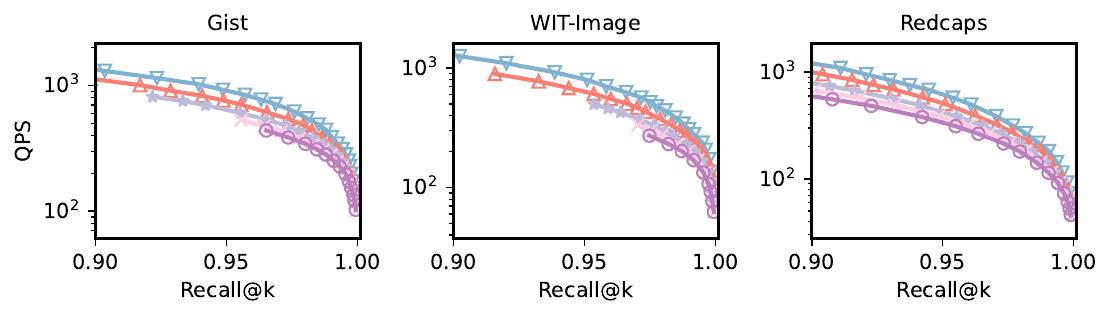}
  \vspace{-5mm}
  \caption{Impact of $k$ values (Exp. 11)}
  \label{fig:interval_vary_k}
  \vspace{-3mm}
\end{figure}

\noindent \textbf{Exp. 11: Impact of $k$.}
Fig.~\ref{fig:interval_vary_k} presents the impact of $k$ on the performance of MSTG. 
As $k$ increases from 10 to 100, a gradual decrease in RRANN search performance is observed, due to the increased number of verified candidates for the same accuracy.

\begin{figure*}[t]
  \centering
   \includegraphics[width=0.23\linewidth]{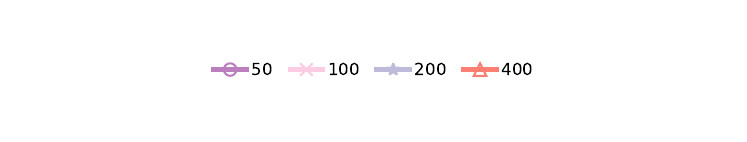}
  \includegraphics[width=\linewidth]{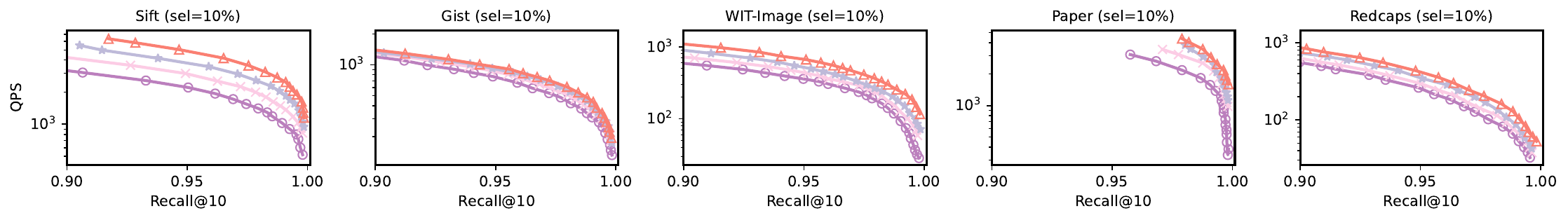}
  \vspace{-6mm}
  \caption{Impact of $ef_{con}$ values (Exp. 12)}
  \label{fig:interval_vary_efcon}
  \vspace{-2mm}
\end{figure*}

\noindent \textbf{Exp. 12: Impact of Parameter $ef_{con}$.}
Fig.~\ref{fig:interval_vary_efcon} shows the impact of $ef_{con}$ on search with query selectivity as 5\% on Gist.
We can see that increasing $ef_{con}$ slightly improves the search performance, because a larger $ef_{con}$ indicates more candidate neighbors for pruning, which enhances the quality of the built HNSW at the expense of construction cost. 
Moreover, once $ef_{con}$ reaches 200, further increasing $ef_{con}$ yields negligible performance gains. Thus, we set $ef_{con}=200$ by default in our experiments.

\begin{figure*}[t]
  \centering
   \includegraphics[width=0.2\linewidth]{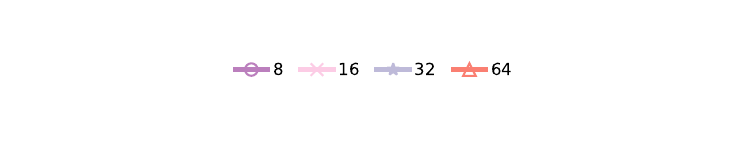}
  \includegraphics[width=\linewidth]{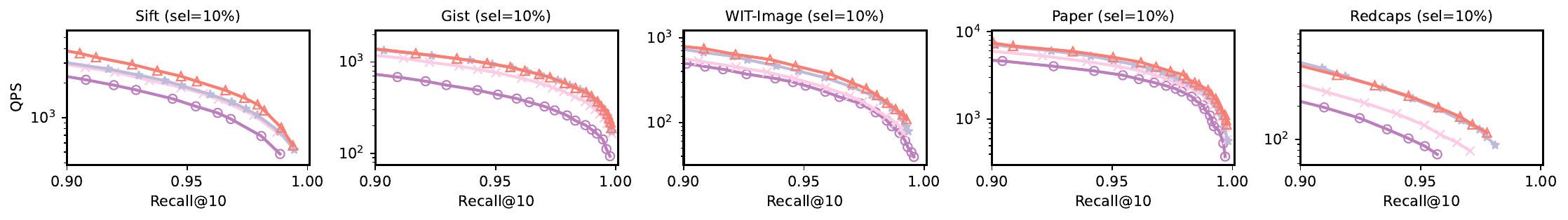}
  \vspace{-6mm}
  \caption{Impact of $M$ values (Exp. 13)}
  \label{fig:interval_vary_m}
  \vspace{-2mm}
\end{figure*}

\noindent \textbf{Exp. 13: Impact of Parameter $M$.}
Fig.~\ref{fig:interval_vary_m} shows the impact of $M$, which controls the upper bound of node out-degree in HNSW.
Increasing $M$ leads to better performance from 8 to 16, where the QPS at Recall@10 = 0.90 nearly doubles, due to improved graph connectivity. However, further increasing $M$ yields only marginal improvements while significantly increasing the building cost. To strike a balance between search performance and construction efficiency, we set $M = 32$ in our experiments.



\end{document}